\documentclass[11pt,a4paper]{article}
\usepackage{a4wide}
\usepackage{amsmath,amssymb}
\usepackage[mathscr]{eucal}
\newtheorem{definition}{Definition}
\newtheorem{theorem}{Theorem}
\newtheorem{lemma}{Lemma}
\newtheorem{proposition}{Proposition}
\newtheorem{corollary}{Corollary}
\begin{document}
\title{Forward and backward time observables for\\
  quantum evolution and\\ 
  quantum stochastic processes -I: The time observables\\}
\author{Y. Strauss\thanks{E-mail: ystrauss@math.bgu.ac.il}\\
            Department of Mathematics\\
            Ben-Gurion University of the negev\\
            Be'er Sheva 84105, Israel}
\date{}
\maketitle
\begin{abstract}
Given a Hamiltonian $\mathbf H$ on a Hilbert space $\mathcal H$ it is shown that, under the assumption that 
$\sigma(\mathbf H)=\sigma_{ac}(\mathbf H)=\mathbb R^+$, there exist unique positive operators $\mathbf T_F$ and 
$\mathbf T_B$ registering the Schr\"odinger time evolution generated by $\mathbf H$ in the forward (future) direction and 
backward (past) direction respectively. These operators may be considered as time observables for the quantum evolution.
Moreover, it is shown that the same operators may serve as time observables in the construction of quantum stochastic
differential equations and quantum stochastic processes in the framework of the Hudson-Parthasarathy quantum stochastic
calculus. The basic mechanism enabling for the definition of the time observables originates from the recently developed semigroup
decomposition formalism used in the description of the time evolution of resonances in quantum mechanical scattering problems.
\end{abstract}
\section{Introduction}
\label{introduction}
\par The recently developed semigroup decomposition formalism for the
description of the time evolution of quantum mechanical resonances \cite{St1,St2,SHV} utilizes two central
ingredients, namely the Sz.-Nagy-Foias theory of contraction operators and
strongly contractive semigroups on Hilbert space \cite{SzNF} and the contractive nesting of
Hilbert spaces, i.e., the embedding of one Hilbert space into another
via a contractive quasi-affine transformation \cite{Gr}, in order to decompose the time evolution of resonances
in standard, non-relativistic, quantum mechanical scattering problems
into a sum of a semigroup part and a non-semigroup part. In this decomposition the semigroup part,
given in terms of a Lax-Phillips type semigroup (see for example \cite{SHV} for the terminology used here), is the
resonance term and the non-semigroup part is called the background
term. The complex eigenvalues of the generator of the semigroup, providing the
typical exponential decay behaviour of the resonance part, are associated with resonance poles of the
scattering matrix. In fact, under appropriate conditions, the
scattering matrix can be factored and the rational part of this
factorization corresponds to the characteristic function (see for example \cite{SzNF}) of
the cogenerator of the (adjoint of) the resonance part semigroup. By a theorem of
Sz.-Nagy and Foias (\cite{SzNF}, Chap. VI, Sec. 4) the characteristic function determines
uniquely the spectrum of the generator of the resonant part
Lax-Phillips type semigroup. The close relation between resonance poles
of the S-matrix and the time evolution of a resonance is thus clearly
exhibited.
\par As mentioned above, the semigroup decomposition formalism uses
the Sz.-Nagy-Foias theory in order to extract the semigroup part of
the evolution of a resonance. Specifically, use is made of a Hardy
space functional model for the $C_{\cdot 0}$ class
contractive semigroup (a Lax-Phillips type semigroup in the
terminology used above) corresponding to the exponentially decaying
resonance part of the evolution. This functional model
is associated with the construction of isometric dilations of
$C_{\cdot 0}$ class semigroups \cite{SzNF, K, H, LP, Rot,Rov} (see also \cite{St1} for a short
review of the mathematical structures involved). A central ingredient of this
functional model is a particular semigroup on Hardy space
which we will presently define.
\par Denote by $\mathbb C^+$ the upper
half of the complex plane and let $\mathscr H^2_{\mathcal N}(\mathbb C^+)$ be the
Hardy space of vector valued functions analytic in the upper
half-plane and taking values in a separable Hilbert space $\mathcal N$. Similarly, 
$\mathscr H^2_{\mathcal N}(\mathbb C^-)$ denotes the Hardy space of $\mathcal N$ valued functions 
analytic in the lower half-plane $\mathbb C^-$. The set of boundary values on
$\mathbb R$ of functions in $\mathscr H^2_{\mathcal N}(\mathbb C^+)$ is a Hilbert
space isomorphic to $\mathscr H^2_{\mathcal N}(\mathbb C^+)$ which we denote by
$\mathscr H^+_{\mathcal N}(\mathbb R)$. In a similar manner $\mathscr H^-_{\mathcal N}(\mathbb R)$,
the space of boundary values of functions in $\mathscr H^2_{\mathcal N}(\mathbb C^-)$, is isomorphic to 
$\mathscr H^2_{\mathcal N}(\mathbb C^-)$. Throughout the present paper we mostly deal with $\dim\mathcal N=1$, i.e.,
the case scalar valued functions. In this case we denote by $\mathscr H^2(\mathbb C^+)$ and $\mathscr H^2(\mathbb C^-)$
the Hardy spaces of the upper half-plane and lower half-plane respectively. The spaces of boundary value functions on the real axis
are then denoted by $\mathscr H^+(\mathbb R)$ and $\mathscr H^-(\mathbb R)$. Define a family
$\{u(t)\}_{t\in\mathbb R}$ of unitary, multiplicative operators
$u(t):L^2_{\mathcal N}(\mathbb R)\mapsto L^2_{\mathcal N}(\mathbb R)$
by
\begin{equation}
\label{translation_group_eqn}
 [u(t)f](\sigma)=e^{-i\sigma t}f(\sigma),\quad f\in L^2_{\mathcal
   N}(\mathbb R),\quad \sigma\in\mathbb R\,,
\end{equation}
The family $\{u(t)\}_{t\in\mathbb R}$ forms a one parameter group of multiplicative operators on 
$L^2_{\mathcal N}(\mathbb R)$. Let $P_+$ be the orthogonal projection of $L^2_{\mathcal N}(\mathbb
R)$ on $\mathscr H^+_{\mathcal N}(\mathbb R)$. A \emph{Toeplitz operator} with
symbol $u(t)$ \cite{N1,N2, RosR} is an operator $T^+_u(t):\mathscr
H^+_{\mathcal N}(\mathbb R)\mapsto \mathscr H^+_{\mathcal N}(\mathbb R)$ defined by
\begin{equation}
\label{toeplitz_semigroup_eqn}
 T^+_u(t)f:=P_+u(t)f,\qquad f\in\mathscr H^+_{\mathcal N}(\mathbb R)\,.
\end{equation}
The set $\{T^+_u(t)\}_{t\in\mathbb R^+}$ forms a strongly
 continuous, contractive, one parameter semigroup on $\mathscr H^+_{\mathcal N}(\mathbb R)$ satisfying 
\begin{equation*}
 s-\lim_{t\to\infty} T^+_u(t) =0\,.
\end{equation*}
Moreover, we have
\begin{equation*}
 \Vert T^+_u(t_2)f\Vert\leq\Vert T^+_u(t_1)f\Vert,\qquad t_2>t_1,\
 f\in\mathscr H^+_{\mathcal N}(\mathbb R)\,.
\end{equation*}
The functional model providing the semigroup evolution of the
resonance term in the semigroup decomposition formalism is obtained by
the compression of the semigroup $\{T^+_u(t)\}_{t\in\mathbb
  R^+}$ to an invariant subspace $\hat{\mathscr K}\subset\mathscr H^+_{\mathcal
  N}(\mathbb R)$ given by 
\begin{equation}
\label{compress_subspace_eqn}
 \hat{\mathscr K}=\mathscr H^+_{\mathcal N}(\mathbb
 R)\ominus\Theta_T(\cdot)\mathscr H^+_{\mathcal N}(\mathbb R)
\end{equation}
where $\Theta_T(\cdot):\mathscr H^+_{\mathcal N}(\mathbb R)\mapsto
\mathscr H^+_{\mathcal N}(\mathbb R)$ is an \emph{inner function} \cite{RosR,N1} (also
\cite{Hof,D} for the scalar case)
for $\mathscr H^+_{\mathcal N}(\mathbb R)$.
In the semigroup decomposition formalism $\Theta_T(\cdot)$ 
is associated with the scattering matrix. In fact, it corresponds to the rational factor in
the factorization of the $S$-matrix mentioned above. 
\par Observe that the semigroups above are defined in terms of
abstract function spaces i.e., Hardy spaces and certain subspaces of
Hardy spaces. A natural question is how are these semigroups linked to
the physical Hilbert space and physical evolution of a given quantum
mechanical system. In the semigroup decomposition formalism this is the role of the quasi-affine transform
mentioned above. We first recall the definition of a quasi-affine map
(in Definition \ref{quasi_affine_map_def} and through the rest of the paper $\mathcal B(\mathcal H)$
denotes the space of bounded linear operators on $\mathcal H$)
\begin{definition}[quasi-affine map]
\label{quasi_affine_map_def}
A quasi-affine map from a Hilbert space $\mathcal H_1$ into a Hilbert
space $\mathcal H_0$ is a linear, injective, continuous mapping of
$\mathcal H_1$ into a dense linear manifold in $\mathcal H_0$. If
$\mathbf A\in\mathcal B(\mathcal H_1)$ and $\mathbf B\in \mathcal
B(\mathcal H_0)$ then $\mathbf A$ is a quasi-affine transform of
$\mathbf B$ if there is a quasi-affine map $\theta:\mathcal
H_1\mapsto\mathcal H_0$ such that $\theta\mathbf A=\mathbf B\theta$.
\end{definition}
If $\theta: \mathcal H_1\mapsto\mathcal H_0$ is a
quasi-affine mapping then $\theta^*: \mathcal
H_0\mapsto\mathcal H_1$ is also quasi-affine,
that is $\theta^*$ is one to one, continuous and its range is dense 
in $\mathcal H_1$. Moreover, if $\theta_1: \mathcal
H_0\mapsto\mathcal H_1$ is quasi-affine and $\theta_2:\mathcal
H_1\mapsto\mathcal H_2$ is quasi-affine then
$\theta_2\theta_1:\mathcal H_0\mapsto\mathcal H_2$ is quasi-affine \cite{SzNF}. 
\par Consider a seperable Hilbert space $\mathcal H$ and a one
parameter unitary evolution group $\{\mathbf U(t)\}_{t\in\mathbb R}$
on $\mathcal H$ generated by a self-adjoint Hamiltonian $\mathbf H$.
We will assume that the spectrum of $\mathbf H$ satisfies
$\sigma(\mathbf H)=\sigma_{ac}(\mathbf H)=ess\,supp\,\,\sigma_{ac}(\mathbf H)=\mathbb R^+$. For simplicity we will assume
furthermore that the multiplicity of $\sigma(\mathbf H)$ is one. By a slight
variation of a fundamental theorem proved in reference \cite{St1} one can then prove the existence of a mapping $\Omega_f:
\mathcal H\mapsto \mathscr H^+(\mathbb R)$ such that 
\begin{itemize}
\item[(i)] $\Omega_f$ is a contractive quasi-affine mapping of
      $\mathcal H$ into $\mathscr H^+(\mathbb R)$.
\item[(ii)] For $t\geq 0$, $\mathbf U(t)$ is a quasi-affine transform
      of the Toeplitz operator $T^+_u(t)$. For every $g\in\mathcal H$ we have
\begin{equation}
\label{basic_intertwining_rel_eqn}
 \Omega_f\mathbf U(t)g=T^+_u(t)\Omega_f g,\qquad t\geq 0\,.
\end{equation}
\end{itemize}
(here the subscript $f$ in $\Omega_f$ designates forward time evolution).
We note that in the case of the semigroup decomposition one constructs
two different such quasi-affine transformations, denoted
$\hat\Omega_\pm$ and corresponding respectively to the two M\o ller wave
operators for a given scattering problem. One then defines what is
called the nested $S$-matrix $S_{nest}:=\hat\Omega_+\hat\Omega_-^{-1}$
(see appropriate definitions in {\cite{St2}).
\par By the remarks following the definition above of a quasi-affine
mapping, if $\Omega_f:\mathcal H\mapsto\mathscr
H^+(\mathbb R)$ is quasi-affine then $\Omega_f^*: \mathscr H^+(\mathbb
R)\mapsto\mathcal H$ is quasi-affine with range dense
in $\mathcal H$. Denoting $\mathbf T^{-1}_F:=\Omega_f^*\Omega_f$
we find that $\mathbf T^{-1}_F: \mathcal H\mapsto\mathcal H$ is a quasi-affine mapping with range dense in
$\mathcal H$. Setting
$\Sigma_{\Omega_f}:=\text{Ran}\,(\mathbf T_F^{-1})$ one can show that $\mathbf T^{-1}_F$ is a positive,
bounded, self-adjoint operator with a self-adjoint inverse $\mathbf T_F: \Sigma_{\Omega_f}\mapsto\mathcal H$
(see Section \ref{the_time_observables_sec}). We have
\begin{definition}[forward time observables]
Let $\Omega_f:\mathcal H\mapsto\mathscr H^+(\mathbb
R)$ be a mapping satisfying (i)-(ii) above, and let
$\Sigma_{\Omega_f}=\Omega_f^*\Omega_f\,\mathcal H$ and $\hat\Sigma_{\Omega_f}=\Omega_f\Omega_f^*\mathscr H^+(\mathbb R)$. Then the operator $\hat T_F:
\hat\Sigma_{\Omega_f}\mapsto\mathcal H^+(\mathbb R)$ defined by
\begin{equation*}
 \hat T_F:= (\Omega_f^*)^{-1}\Omega_f^{-1}
\end{equation*}
will be called the Hardy space forward time observable and the
operator $\mathbf T_F: \Sigma_{\Omega_f}\mapsto \mathcal H$
defined by
\begin{equation*}
 \mathbf T_F:=\Omega_f^{-1}(\Omega_f^*)^{-1}
\end{equation*}
will be called the physical forward time observable.\hfill$\square$
\end{definition}
It is proved in Section \ref{the_time_observables_sec} that $\inf\sigma(\mathbf T_F)=1$ 
where $\sigma(\mathbf T_F)$ denotes the spectrum of $\mathbf T_F$.
An extended version of the following theorem is also proved in Section \ref{the_time_observables_sec}:
\begin{theorem}
\label{forward_t_observable_thm}
Let $\xi_{\mathcal H}$ be the spectral measure (i.e. spectral projection valued measure) of $\mathbf T_F$ 
and let $a>1$. Then for any $g\in\mathcal H$ satisfying $\xi_{\mathcal H}([1,a))g=g$ there exists $\tau>0$ such
that $\xi_{\mathcal H}([a,\infty))\mathbf U(t)g\not=\{0\}$ for all $t>\tau$ and, moreover, $\lim_{t\to\infty}\Vert\xi_{\mathcal
  H}([1,a))\mathbf U(t)g\Vert_{\mathcal H}=0$.
\par\hfill$\square$
\end{theorem}
Theorem \ref{forward_t_observable_thm} states that if $g\in\mathcal H$ is
compactly supported on the spectrum of $\mathbf T_F$  then the evolved state
$g(t)=\mathbf U(t)g$ in the forward direction of time (i.e., for
$t\geq 0$) must ``go up'' on the spectrum of $\mathbf T_F$
as time increases. Therefore, \emph{a priori} $\mathbf T_F$ may be regarded as an observable
registering the flow of time in the system in the future direction. This observation provides the
motivation for the terminology used for $\mathbf T_F$ in the definition above.
Note that $\mathbf T_F$ is not a time observable in the sense of a Mackey
imprimitivity system \cite{M} (such a time
observable does not exist for problems where the generator of evolution
is semibounded).
\par Although formulated completely within the framework of standard
quantum mechanics, where the evolution of a system is given in terms
of a one parameter group $\{\mathbf U(t)\}_{t\in\mathbb R}$ generated
by a self-adjoint Hamiltonian $\mathbf H$, the semigroup decomposition
formalism holds its origines in a particular line of investigation associated with recent efforts to
understand irreversible quantum evolution (in this respect see \cite{P,EH,HoP,St1,StHE,Ba1,Ba2}). The time asymmetry built
into this framework is clearly exhibited in property (ii) above where
the intertwining of $\mathbf U(t)$ and $T^+_u(t)$ through $\Omega_f$
is valid only for $t\geq 0$. In fact, one may apply the semigroup decomposition
also in the backward direction of time using the lower half-plane
Hardy space $\mathscr H^-(\mathbb R)$ and a different quasi-affine
mapping. Denoting by $P_-$ the orthogonal projection of $L^2(\mathbb R)$ on $\mathscr H^-(\mathbb R)$ 
we consider in $\mathscr H^-(\mathbb R)$ the family of operators $T^-_u(t):\mathscr H^-(\mathbb R)
\mapsto\mathscr H^-(\mathbb R)$ defined by
\begin{equation*}
 T^-_u(t)f:=P_-u(t)f, \qquad  f\in\mathscr H^-(\mathbb R)\,.
\end{equation*}
The set $\{T^-_u(-t)\}_{t\in\mathbb R^+}$ forms a strongly continuous, contractive, one parameter semigroup on 
$\mathscr H^-(\mathbb R)$. Then, under the same assumptions leading to (i) and (ii) above, there exists a transformation 
$\Omega_b: \mathcal H\mapsto \mathscr H^-(\mathbb R)$ with the properties
\begin{itemize}
\item[(i')] $\Omega_b$ is a contractive quasi-affine mapping of $\mathcal H$ into $\mathscr H^-(\mathbb R)$.
\item[(ii')] For $t\leq 0$, $\mathbf U(t)$ is a quasi-affine transform of $T^-_u(t)$. 
For every $f\in\mathcal H$ we have
\begin{equation}
\label{back_basic_intertwine_eqn}
 \Omega_b\mathbf U(t)f=T^-_u(t)\Omega_b f,\qquad t\leq 0\,.
\end{equation}
\end{itemize}
(here the subscript $b$ in $\Omega_b$ designates backward time evolution).
Using Eq. (\ref{basic_intertwining_rel_eqn}) and Eq. (\ref{back_basic_intertwine_eqn})
in the semigroup decomposition formalism the description of the evolution of a system may be splits into two
different representations, one in the forward direction and one in the
backward direction, according to the different semigroup acting in
each of them. The structure thus obtained resembles, and is in fact closely related
to, the use of Hardy spaces in a rigged Hilbert space formulation of
the problem of resonances \cite{Bh1,Bh2,BhG,BhK, Ga1,Ga2} where the evolution also splits
into a semigroup acting in the forward direction and a differnt
semigroup acting in the backward direction. 
\par Associated with the quasi-affine mapping $\Omega_b$ are the
Hardy space and physical space backward time observables
\begin{definition}[backward time observables]
Let $\Omega_b:\mathcal H\mapsto\mathscr H^-(\mathbb
R)$ be a map satisfying (i')-(ii') above and let
$\Sigma_{\Omega_b}:=\Omega_b^*\Omega_b\,\mathcal H$ and $\hat\Sigma_{\Omega_b}:=\Omega_b\Omega_b^*\mathscr H^-(\mathbb R)$. Then the operator $\hat T_B:
\hat\Sigma_{\Omega_b}\mapsto\mathcal H^-(\mathbb R)$ defined by
\begin{equation*}
 \hat T_B:= (\Omega_b^*)^{-1}\Omega_b^{-1}
\end{equation*}
will be called the Hardy space backward time observable and the
operator $\mathbf T_B: \Sigma_{\Omega_b}\mapsto \mathcal H$
defined by
\begin{equation*}
 \mathbf T_B:=\Omega_b^{-1}(\Omega_b^*)^{-1}
\end{equation*}
will be called the physical backward time observable.\hfill$\square$
\end{definition}
A Theorem analogous to Theorem \ref{forward_t_observable_thm} holds for $\mathbf T_B$ for negative times. The reader is referred 
to Theorem \ref{time_observable_thm} in Section \ref{the_time_observables_sec}.
\par Considering the existence of distinct time observables for forward
and backward evolution and the \emph{a priori} time asymmetry existing in the
formalism from which they arise, one may naturally ask whether these operators
can be used as time observables not only for a quantum
system undergoing Schr\"odinger evolution but more generally for quantum
irreversible processes such as quantum stochastic processes. Another
question is whether relations similar to those in (ii) and (ii') above between
evolution in Hardy space and evolution in physical space
are again exhibited in this more general context. In Section \ref{t_observe_stochastic_process_sec} below
we consider the time observables $\mathbf T_F$, $\hat T_F$ in the framework of the
Hudson-Pharthasarathy (HP) quantum stochastic calculus \cite{HuP,Par1,Par2} and show that
indeed these operators can be used as time
observables for quantum stochastic processes and that the stochastic processes defined with respect to the (second quantisation
of) Hardy space can be mapped to corresponding stochastic processes
defined with respect to the (second quantisation of) physical space 
through a mapping associated with the quasi-affine map
$\Omega_f:\mathcal H\mapsto\mathscr H^+(\mathbb R)$.
\par Some remarks concerning notation: With the exception of the identity, operators acting in $\mathcal H$ are
denoted below by capital bold face letters. Thus $\mathbf H$, $\mathbf K$, $\mathbf T_F$ etc. are all operators in 
$\mathcal H$. Operators acting in the Hardy spaces $\mathscr H^+(\mathbb R)$ and $\mathscr H^-(\mathbb R)$
are denoted by a hat. Thus $\hat H$, $\hat K$, $\hat T_F$ etc. are operators in Hardy spaces. Unless otherwise specifically 
stated in the text the identity operators in the various Hilbert spaces are generically denote simply by $1$ with the exact meaning
implied by the particular context. The Borel $\sigma$-algebra
of $\mathbb R^+$ is denoted by $\mathfrak B^+$ and the set of all projection operators in a Hilbert space $\mathcal H$ is
denoted $\mathcal P(\mathcal H)$ (thus, for example, to a positive self-adjoint operator $\mathbf A$ there corresponds by the
spectral theorem a spectral projection valued measure, say $\xi$, such that $\xi:\mathfrak B^+\mapsto\mathcal P(\mathcal H)$).
In addition the norm in the various Hilbert spaces is denoted by the appropriate subscript; in particular 
$\Vert\cdot\Vert_{\mathscr H^+}$ denotes the norm in $\mathscr H^+(\mathbb R)$,
$\Vert\cdot\Vert_{\mathscr H^-}$ denotes the norm in $\mathscr H^-(\mathbb R)$ etc.\ .
\par The rest of this paper is organized as follows: In Subsection \ref{def_basic_properties_subsec} an exact definition of
forward and backward time observables is given followed by a discussion of several of their basic properties such as domain of
definition, positivity, self-adjointness etc.\ . Subsequently, Theorem \ref{time_observable_thm} in the same subsection establishes
the motivation for the terminology "physical time observables" applied to the operators $\mathbf T_F$ and $\mathbf T_B$. This
theorem is of central importance in the context of the present work. Section \ref{the_time_observables_sec} concludes in
Subsection \ref{origin_time_observe_subsec} with a discussion of the mathematical structure enabling the existence of time
observables. The application of the operators $\mathbf T_F$, $\hat T_F$ (and $\mathbf T_B$, $\hat T_B$) as time observables
for quantum stochastic processes is discussed in Section \ref{t_observe_stochastic_process_sec}. The goal is to find the
analogue of the fundamental intertwining relation in Eq. (\ref{basic_intertwining_rel_eqn}) in the context of quantum stochastic
processes. This is done in two steps; first Subsection \ref{creation_annihilation_conservation_map_subsec} deals with the
mapping of basic creation, annihilation and conservation processes defined with respect to $\mathbf T_F$ into
the corresponding processes defined with respect to $\hat T_F$, then Subsection \ref{stoch_process_map_subsection} contains
an application of the mapping defined in the previous subsection to an important class of quantum stochastic differential equations
and an analogue of Eq. (\ref{basic_intertwining_rel_eqn}) is obtained for this class. Section \ref{summary_sec} contains a
summary of results and open questions.
\section{The time observables}
\label{the_time_observables_sec}
\subsection{Definition and basic properties}
\label{def_basic_properties_subsec}
\par We first remark that statements concerning forward time observables $\mathbf T_F$ and $\hat T_F$
and backward time observables $\mathbf T_B$ and $\hat T_B$ are proved essentially using the same methods with obvious 
necessary changes. Therefore for the sake of completeness theorems are stated with specific reference to forward and 
backward time observables whereas detailed proofs are given for the forward time observables with an indication of the 
necessary replacements pertaining to backward time observables.
\par As in Section \ref{introduction} above let $\mathbf T^{-1}_F: \mathcal H\mapsto\mathcal H$ be defined by $\mathbf T^{-1}_F:=\Omega_f^*\Omega_f$ and
$\hat T^{-1}_F:\mathscr H^+(\mathbb R)\mapsto\mathscr H^+(\mathbb R)$
be defined by $\hat T_F^{-1}:=\Omega_f\,\Omega_f^*$. Let 
$\Sigma_{\Omega_f}=\text{Ran}\,\mathbf T_F^{-1}=\Omega_f^*\Omega_f\,\mathcal H$ 
and $\hat\Sigma_{\Omega_f}=\text{Ran}\,\hat
T_F^{-1}=\Omega_f\,\Omega_f^*\,\mathscr H^+(\mathbb R)$. Let $\mathbf T_B^{-1}:\mathcal H\mapsto
\mathcal H$ be defined by $\mathbf T_B^{-1}:=\Omega_b^* \Omega_b$ and $\hat T_B^{-1}:\mathscr H^-(\mathbb R)
\mapsto\mathscr H^-(\mathbb R)$ be defined by $\hat T_B^{-1}:=\Omega_b \Omega_b^*$. Let $\Sigma_{\Omega_b}=\text{Ran}\,
\mathbf T_B^{-1}=\Omega_b^* \Omega_b\mathcal H$ and $\hat\Sigma_{\Omega_b}=\text{Ran}\,\hat T_B^{-1}
=\Omega_b \Omega_b^*\,\mathscr H^-(\mathbb R)$. We have
\begin{proposition}
\label{time_operators_prop}
The operator $\mathbf T_F: \mathcal D(\mathbf T_f)\mapsto \mathcal H$
given by $\mathbf T_F:=(\mathbf T_F^{-1})^{-1}=\Omega_f^{-1}(\Omega_f^*)^{-1}$ is
a positive, unbounded self-adjoint operator in $\mathcal H$ with domain 
$\mathcal D(\mathbf T_F)=\Sigma_{\Omega_f}$. Similarly, the operator $\hat T_F:
\mathcal D(\hat T_F)\mapsto\mathscr H^+(\mathbb R)$ given by
$\hat T_F:=(\hat T_F^{-1})^{-1}=(\Omega_f^*)^{-1}\Omega_f^{-1}$ is a positive,
unbounded self-adjoint operator in $\mathscr H^+(\mathbb R)$ with domain $\mathcal D(\hat T_f)=\hat\Sigma_{\Omega_f}$ . The
spectrums $\sigma(\mathbf T_F)$ and $\sigma(\hat T_F)$ satisfy
$\sigma(\mathbf T_F)=\sigma(\hat T_F)$. If $\xi_{\mathscr H^+}: \mathfrak B^+\mapsto \mathcal P(\mathscr H^+(\mathbb R))$ is the spectral 
measure (i.e., spectral projection valued measure) of $\hat T_F$ and $\xi_{\mathcal H}:\mathfrak B^+\mapsto\mathcal P(\mathcal H)$ 
is the spectral measure of $\mathbf T_F$ then, for any set $E\in\mathfrak B^+$ we have 
\begin{equation}
\label{spectral_family_transform_A_eqn}
 \xi_{\mathcal H}(E)g=\Omega_f^{-1}\xi_{\mathscr H^+}(E)\Omega_f g,\qquad g\in\text{Ran}\,\Omega_f^*\,,
\end{equation}
\begin{equation}
\label{spectral_family_transform_B_eqn}
 \xi_{\mathscr H^+}(E)g=(\Omega^*_f)^{-1}\xi_{\mathcal H}(E)\Omega^*_f g,\qquad g\in\text{Ran}\,\Omega_f\,,
\end{equation}
so that
\begin{equation*}
 \xi_{\mathcal H}(E)=\overline{\Omega_f^{-1}\xi_{\mathscr H^+}(E)\Omega_f},\qquad
 \xi_{\mathscr H^+}(E)=\overline{(\Omega^*_f)^{-1}\xi_{\mathcal H}(E)\Omega^*_f}\,.
\end{equation*}
The operator $\mathbf T_B: \mathcal D(\mathbf T_B)\mapsto \mathcal H$
given by $\mathbf T_B:=(\mathbf T_B^{-1})^{-1}=\Omega_b^{-1}(\Omega_b^*)^{-1}$ is
a positive, unbounded self-adjoint operator in $\mathcal H$ with domain 
$\mathcal D(\mathbf T_B)=\Sigma_{\Omega_b}$. Similarly, the operator $\hat T_B:
\mathcal D(\hat T_F)\mapsto\mathscr H^-(\mathbb R)$ given by
$\hat T_B:=(\hat T_B^{-1})^{-1}=(\Omega_b^*)^{-1}\Omega_b^{-1}$ is a positive,
unbounded self-adjoint operator in $\mathscr H^-(\mathbb R)$ with domain $\mathcal D(\hat T_F)=\hat\Sigma_{\Omega_b}$. The
spectrums $\sigma(\mathbf T_B)$ and $\sigma(\hat T_B)$ satisfy
$\sigma(\mathbf T_B)=\sigma(\hat T_B)$. If $\zeta_{\mathscr H^-}: \mathfrak B^+\mapsto\mathcal P(\mathscr H^-(\mathbb R))$ is the spectral 
measure of $\hat T_B$ then, for any set $E\in\mathfrak B^+$ we have 
\begin{equation}
\label{back_spectral_family_transform_A_eqn}
 \zeta_{\mathcal H}(E)g=\Omega_b^{-1}\zeta_{\mathscr H^-}(E)\Omega_b g,\qquad g\in\text{Ran}\,\Omega_b^*\,,
\end{equation}
\begin{equation}
\label{back_spectral_family_transform_B_eqn}
 \zeta_{\mathscr H^-}(E)g=(\Omega_b^*)^{-1}\zeta_{\mathcal H}(E)\Omega^*_bg,\qquad g\in\text{Ran}\,\Omega_b\,,
\end{equation}
so that
\begin{equation*}
 \zeta_{\mathcal H}(E)=\overline{\Omega_b^{-1}\zeta_{\mathscr H^-}(E)\Omega_b},\qquad
 \zeta_{\mathscr H^-}(E)=\overline{(\Omega^*_b)^{-1}\zeta_{\mathcal H}(E)\Omega^*_b}\,.
\end{equation*}
\hfill$\square$
\end{proposition}
\par{\bf Proof of Proposition \ref{time_operators_prop}:}
\par Clearly the operators $\mathbf T_F^{-1}$ and $\hat T_F^{-1}$ are
positive and symmetric. Since both $\Omega_f$ and
$\Omega_f^*$ are quasi-affine maps we have
$\text{Ker}\,\mathbf T_F=\{0\}$, $\text{Ker}\,\hat T_F=\{0\}$ and, since both
are also contractive we have $\Vert\mathbf T_F\Vert\leq 1$, $\Vert\hat
T_F\Vert\leq 1$ so that, in particular, $\text{Dom}\,
\mathbf T_F^{-1}=\mathcal H$ and $\text{Dom}\,\hat T_F^{-1}=\mathscr
H^+(\mathbb R)$. We conclude that $T_F^{-1}$ and $\hat T_F^{-1}$ are
self-adjoint. Moreover, again by the fact that both $\Omega_f$
and $\Omega_f^*$ are quasi-affine, we have that
$\text{Ran}\,\mathbf T_F^{-1}=\Sigma_{\Omega_f}\subset\mathcal H$ is dense in $\mathcal H$ and
$\text{Ran}\,\hat T_F^{-1}=\hat\Sigma_{\Omega_f}\subset\mathscr
H^+(\mathbb R)$ is dense in $\mathscr H^+(\mathbb R)$. Therefore,
$\mathbf T_F^{-1}$ and $\hat T_F^{-1}$ are invertible on a dense domain and so
(see for example \cite{RSzN}) $\mathbf T_F=(\mathbf T_F^{-1})^{-1}$ and $\hat
T_F=(\hat T_F^{-1})^{-1}$ are self-adjoint with domains 
$\text{Dom}\,\mathbf T_F=\Sigma_{\Omega_f}$ and $\text{Dom}\,\hat T_F=\hat\Sigma_{\Omega_f}$ respectively.
\par The operators $\mathbf T_F$ and $\hat T_F$ cannot be extended to bounded operators.
consider the two maps $\Omega_f: \mathcal H\mapsto\mathscr H^+(\mathbb R)$ and $\Omega_f^*: \mathscr H^+(\mathbb R)\mapsto\mathcal H$. Since both maps are quasi-affine we know that
$\text{Ran}\,\Omega_f$ is dense in $\mathscr H^+(\mathbb R)$ and $\text{Ran}\,\Omega_f^*$ is dense in 
$\mathcal H$. By the injective property of quasi-affine maps the two maps are also invertible on their range. 
Furthermore, we have (see \cite{RSzN})
\begin{equation}
\label{inv_star_exchange_eqn}
 (\Omega_f^{-1})^*=(\Omega_f^*)^{-1}\,.
\end{equation}
It follows from Eq. (\ref{inv_star_exchange_eqn}) that $\Omega_f^{-1}$ and $(\Omega_f^*)^{-1}$ cannot be bounded. For 
$(\Omega_f^*)^{-1}: \Omega_f^*\mathscr H^+(\mathbb R)\mapsto\mathcal H$ is onto and if
$(\Omega_f^*)^{-1}$ is bounded on $\text{Ran}\,\Omega_f^*$ we can extend it uniquely to a bounded map defined on all of
$\mathcal H$ which we again denote by $(\Omega_f^*)^{-1}$. Then this extended map must have a non-trivial
kernel. However, assuming that $f\in\text{Ker}\,(\Omega_f^*)^{-1}$, for arbitrary $g\in\mathcal H$ we have
\begin{equation*}
 0=((\Omega_f^*)^{-1}f,\Omega_f g)_{\mathscr H^+(\mathbb R)}=((\Omega_f^{-1})^*f,\Omega_f g)_{\mathscr H^+(\mathbb R)}
 =(f,g)_{\mathcal H}
\end{equation*}
which is impossible unless $f=0$. We conclude that $(\Omega_f^*)^{-1}$ is unbounded. 
Similarly, $\Omega_f^{-1}: \Omega_f\mathcal H\mapsto\mathscr H^+(\mathbb R)$ is onto and 
if it can be extended to a bounded map, again denoted by $\Omega_f^{-1}$, defined on all of $\mathscr H^+(\mathbb R)$ 
then the extended map must have a non-trivial kernel.
However, if $f\in\text{Ker}\,\Omega_f^{-1}$ then, for arbitrary $g\in\mathscr H^+(\mathbb R)$ we have
\begin{equation*}
 0=(\Omega_f^{-1}f,\Omega_f^*g)_{\mathcal H}
 =(((\Omega_f^*)^{-1})^*f,\Omega_f^* g)_{\mathcal H}=(f,g)_{\mathscr H^+(\mathbb R)}
\end{equation*}
which is impossible by the arbitreriness of $g$ unless $f=0$. Now,
since $\Omega_f^{-1}$ and $(\Omega_f^*)^{-1}$ are unbounded then
$\mathbf T_F$ and $\hat T_F$ are necessarily unbounded. We note that the fact
that $\Omega_f^{-1}$ and $(\Omega_f^*)^{-1}$ are unbounded implies
that $\inf\sigma(\mathbf T_F^{-1})=\inf\sigma(\hat T_F^{-1})=0$.
\par Let $R_{\mathbf T_F^{-1}}(z)=(z-\mathbf T_F^{-1})^{-1}$ and $R_{\hat T_F^{-1}}(z)=(z-\hat
T_F^{-1})^{-1}$ be, respectively, the resolvents of $\mathbf T_F^{-1}$ and $\hat T_F^{-1}$.
Using the identities $(z-\mathbf T_F^{-1})R_{\mathbf T_F^{-1}}(z)=1$, $(z-\hat T_F^{-1})R_{\hat T_F^{-1}}(z)=1$ 
and $\Omega_f(z-\mathbf T_F^{-1})=(z-\hat T_F^{-1})\Omega_f$
and $\Omega_f^*(z-\hat T_F^{-1})=(z-\mathbf T_F^{-1})\Omega_f^*$ it is easy to
verify that 
\begin{eqnarray}
 R_{\mathbf T_F^{-1}}(z)&=&z^{-1}\left(\Omega_f^*R_{\hat T_F^{-1}}(z)\Omega_f+1\right)\label{rtf_hat_rtf_eqn}\\
 R_{\hat T_F^{-1}}(z)&=&z^{-1}\left(\Omega_f R_{\mathbf T_F^{-1}}(z)\Omega_f^*+1\right)\,.\label{hat_rtf_rtf_eqn}
\end{eqnarray}
Eqns. (\ref{rtf_hat_rtf_eqn}), (\ref{hat_rtf_rtf_eqn}) imply the equality of spectrum
$\sigma(\mathbf T_F^{-1})=\sigma(\hat T_F^{-1})$. Inverting $\mathbf T_F^{-1}$ and $\hat T_F^{-1}$ it follows that 
$\sigma(\mathbf T_F)=\sigma(\hat T_F)$. We note that, since $\mathbf T_F^{-1}$ and $\hat T_F^{-1}$ are contractive, 
one can also utilize the theory of contraction operators on Hilbert space, and especially the notion of 
characteristic functions, to prove the equality of the spectrums. 
This point of view is illuminating and is considered in the appendix where another proof of the equality of the spectrums is 
provided.
\par Let $\xi_{\mathscr H^+}$ be the spectral measure of $\hat T_F$ and, for each
$E\in\mathfrak B^+$ let the operator $\xi_{\mathcal H}(E)$ be defined by the right hand side of
Eq. (\ref{spectral_family_transform_A_eqn}). For any $u\in\text{Ran}\,\Omega_f^*$ we have
\begin{multline*}
 \xi^*_{\mathcal H}(E)u=\Omega_f^*\xi_{\mathscr H^+}(E)(\Omega_f^{-1})^*u
 =\Omega_f^*\xi_{\mathscr H^+}(E)(\Omega_f^*)^{-1}\Omega_f^{-1}\Omega_f u
 =\Omega_f^*\xi_{\mathscr H^+}(E)\hat T_F\Omega_f u=\\
 =\Omega_f^*\hat T_F\xi_{\mathscr H^+}(E)\Omega_f u
 =\Omega_f^{-1}\xi_{\mathscr H^+}(E)\Omega_f u=\xi_{\mathcal H}(E)u\,.\\
\end{multline*}
Hence $\xi_{\mathcal H}(E)$ is symmetric on a dense set in $\mathcal H$ for any set $E\in\mathfrak B^+$. 
Furthermore, for any two sets $E_1,E_2\in\mathfrak B^+$ it easy to
see from the definition that we have
\begin{equation}
\label{spec_meas_intersection_eqn}
 \xi_{\mathcal H}(E_1)\xi_{\mathcal H}(E_2)=\xi_{\mathcal H}(E_2\cap I_1)
\end{equation}
and for $E_1\cap E_2=\emptyset$
\begin{equation}
\label{spec_meas_union_eqn}
 \xi_{\mathcal H}(E_1)+\xi_{\mathcal H}(E_2)=\xi_{\mathcal H}(E_1\cup E_2)\,.
\end{equation}
In particular, if we take in Eq. (\ref{spec_meas_intersection_eqn}) $E_1=E_2=E$ we get
\begin{equation*}
 \xi^2_{\mathcal H}(E)=\xi_{\mathcal H}(E)\,.
\end{equation*}
Thus for each $E\in\mathfrak B^+$, $\xi_{\mathcal H}(E)$ is idempotent and
symmetric on the dense set $\text{Ran}\,\Omega_f^*\subset\mathcal H$ and can be 
extended uniquly to an orthogonal projection on
$\mathcal H$. Eq. (\ref{spec_meas_intersection_eqn}) and
Eq. (\ref{spec_meas_union_eqn}) imply that $\xi_{\mathcal H}$
is a spectral projection valued measure of a self-adjoint operator. Now $\text{Dom}\,\mathbf T_F=\text{Ran}\,\mathbf T_F^{-1}
\subset\text{Ran}\,\Omega_f^*$. Then for every $g\in\text{Dom}\,\mathbf T_F$ we have
\begin{equation*}
 \int_{\sigma(\hat T_F)}\lambda\,\xi_{\mathcal H}(d\lambda) g
 =\Omega_f^{-1}\int_{\sigma(\hat T_F)}\lambda\,\xi(d\lambda)\Omega_f g
 =\Omega_f^{-1}\hat T_F\Omega_f g=\Omega_f^{-1}(\Omega_f^*)^{-1} g=\mathbf T_F g
\end{equation*}
hence $\xi_{\mathcal H}$ is the spectral measure of $\mathbf T_F$.
\par To verify Eq. (\ref{spectral_family_transform_B_eqn}) we note first that the right hand side of this equation is well defined on 
the dense set $\text{Ran}\,\Omega_f$. This is so since we have $(\Omega_f^*)^{-1}\xi_{\mathcal H}(E)\Omega_f^*\Omega_f=
(\Omega_f^*)^{-1}\xi_{\mathcal H}(E)\mathbf T_F^{-1}=(\Omega_f^*)^{-1}\mathbf T_F^{-1}\xi_{\mathcal H}(E)
=\Omega_f\xi_{\mathcal H}(E)$. Plugging Eq. (\ref{spectral_family_transform_A_eqn}) into the right hand side of 
Eq. (\ref{spectral_family_transform_B_eqn}) we get on this dense set
\begin{multline*}
 (\Omega_f^*)^{-1}\xi_{\mathcal H}(E)\Omega_f^* f=\\
 =(\Omega_f^*)^{-1}\Omega_f^{-1}\xi_{\mathscr H^+}(E)\Omega_f\Omega_f^* f
 =\hat T_F\xi_{\mathscr H^+}(E)\hat T_F^{-1} f
 =\xi_{\mathscr H^+}(E)f,\quad f\in\text{Ran}\,\Omega_f\,.
\end{multline*}
The proof of the statements in Proposition \ref{time_operators_prop} concerning backward time observables can be obtained
by following the same steps as for the forward time observables with the obvious replecments of $\Omega_f$ by $\Omega_b$,
$\mathbf T_F$ by $\mathbf T_B$ and $\hat T_F$ by $\hat T_B$.
\hfill$\blacksquare$
\par\bigskip\
\par{\bf Remark:} It is useful to note that, with the help of Eq. (\ref{inv_star_exchange_eqn}) 
we can write $\mathbf T_F$, $\hat T_F$ and $\mathbf T_B$, $\hat T_B$ in a form 
exhibiting more clearly their positivity and symmetric nature, i.e.,
\begin{equation*}
 \mathbf T_F=\Omega_f^{-1}(\Omega_f^*)^{-1}=\Omega_f^{-1}(\Omega_f^{-1})^*,\quad
 \hat T_F=(\Omega_f^*)^{-1}\Omega_f^{-1}=(\Omega_f^{-1})^*\Omega_f^{-1}\,.
\end{equation*}
\begin{equation*}
 \mathbf T_B=\Omega_b^{-1}(\Omega_b^*)^{-1}=\Omega_b^{-1}(\Omega_b^{-1})^*,\quad
 \hat T_B=(\Omega_b^*)^{-1}\Omega_b^{-1}=(\Omega_b^{-1})^*\Omega_b^{-1}\,.
\end{equation*}
\par\bigskip\
The origin of the terminology used for $\mathbf T_F$ and $\mathbf T_B$, i.e., our reference to them as time
observables for forward and backward evolution respectively, follows from the next theorem
\begin{theorem}
\label{time_observable_thm}
We have $\inf\sigma(\mathbf T_F)=\inf\sigma(\mathbf T_B)=1$. 
Let $\xi_{\mathcal H}$ be the spectral measure of $\mathbf T_F$
and let $a>1$. Then, for any $g\in\mathcal H$
satisfying $\xi_{\mathcal H}([1,a))g=g$ there exists
a time $\tau>0$ such that $\xi_{\mathcal H}([a,\infty))\mathbf U(t)g\not=\{0\}$ for
all $t>\tau$ and, moreover, $\lim_{t\to\infty}\Vert\xi_{\mathcal H}([1,a))\mathbf U(t)g\Vert=0$.
Let $\zeta_{\mathcal H}$ be the spectral measure of $\mathbf T_B$. Then for any $g\in\mathcal H$
satisfying $\zeta_{\mathcal H}([1,a))g=g$ there exists a time $\tau'<0$ such that $\zeta_{\mathcal H}([a,\infty))\mathbf U(t)g
\not=\{0\}$ for all $t<\tau'$ and, moreover, $\lim_{t\to -\infty}\Vert\zeta_{\mathcal H}([1,a))\mathbf U(t)g\Vert=0$.
\hfill$\square$
\end{theorem}
\par{\bf Proof:}
\smallskip
\par As above full details are given for the case of $\mathbf T_F$ with an indication of changes necessary for the case of $T_B$.
In order to prove the first part of the theorem we need some more information on the structure of the operators $\Omega_f$ and 
$\Omega_b$. Let $U:\mathcal H\mapsto L^2(\mathbb R^+)$ be the unitary mapping of $\mathcal H$ 
onto its spectral representation on the spectrum of $\mathbf H$ (the energy representation for $\mathbf H$). Let $P_{\mathbb R^+}: L^2(\mathbb R)\mapsto L^2(\mathbb R)$ be the orthogonal projection in $L^2(\mathbb R)$ on the subspace of functions
supported on $\mathbb R^+$ and define the inclusion map $I: L^2(\mathbb R^+)\mapsto L^2(\mathbb R)$ by
\begin{equation}
\label{embedding_eqn}
 (If)(\sigma)=\begin{cases}
                     f(\sigma), & \sigma\geq 0\,,\\
                     0, & \sigma<0\,.\\
                  \end{cases}
\end{equation}
Then the inverse $I^{-1}:P_{\mathbb R^+}L^2(\mathbb R)\mapsto L^2(\mathbb R^+)$ is well defined on 
$P_{\mathbb R^+}L^2(\mathbb R)$. Let $\theta:\mathscr H^+(\mathbb R)\mapsto L^2(\mathbb R^+)$ be given by
\begin{equation}
\label{theta_expression_eqn}
 \theta f=I^{-1}P_{\mathbb R^+}f,\qquad f\in\mathscr H^+(\mathbb R)\,.
\end{equation}
By a theorem of Van Winter \cite{VW}, $\theta$ is a contractive quasi-affine transform mapping $\mathscr H^+(\mathbb R)$ into
$L^2(\mathbb R^+)$. The adjoint map $\theta^*: L^2(\mathbb R^+)\mapsto\mathscr H^+(\mathbb R)$ is then a contractive quasi-affine map.
An explicit expression for $\theta^*$ is given by \cite{St1}
\begin{equation*}
 \theta^*f=P_+If,\qquad f\in L^2(\mathbb R^+)\,.
\end{equation*}
It is shown in \cite{St1} that the maps $\Omega_f$, $\Omega_f^*$ are given by
\begin{equation}
\label{open_omega_exp_eqn}
 \Omega_f=\theta^*U,\qquad \Omega_f^*=U^* \theta\,.
\end{equation}
If in Eq. (\ref{theta_expression_eqn}) instead of functions in $\mathscr H^+(\mathbb R)$ we consider functions in 
$\mathscr H^-(\mathbb R)$ we obtain instead of $\theta$ a different contractive quasi-affine map 
$\overline\theta:\mathscr H^-(\mathbb R)\mapsto L^2(\mathbb R^+)$
\begin{equation}
\label{overline_theta_expression_eqn}
 \overline\theta f=I^{-1}P_{\mathbb R^+}f,\qquad f\in\mathscr H^-(\mathbb R)\,.
\end{equation}
Then $\overline\theta^*:L^2(\mathbb R^+)\mapsto\mathscr H^-(\mathbb R)$ is given by $\overline\theta^*f=P_-If$, $f\in L^2(\mathbb R^+)$ 
and, in a way similar to Eq. (\ref{open_omega_exp_eqn}), one obtains
\begin{equation*} 
\label{open_omega_b_exp_eqn}
 \Omega_b=\overline\theta^*U,\qquad \Omega_b^*=U^* \overline\theta\,.
\end{equation*}
In $\mathscr H^+(\mathbb R)$ consider an element $x_\mu$ of the
form $x_\mu(\lambda)=(\lambda-\mu)^{-1}$, $\text{Im}\,\mu<0$ and denote
$\psi_\mu=\Omega_f^*x_\mu$. We have
\begin{equation*}
 \frac{(x_\mu,\hat T_F^{-1}\,x_\mu)_{\mathscr H^+(\mathbb R)}}{\Vert
   x_\mu\Vert^2_{\mathscr H^+(\mathbb R)}}
 =\frac{(x_\mu,\Omega_f\Omega_f^*x_\mu)_{\mathscr H^+(\mathbb
     R)}}{\Vert x_\mu\Vert^2_{\mathscr H^+(\mathbb R)}}
 =\frac{\Vert\psi_\mu\Vert^2_{\mathcal H}}{\Vert
   x_\mu\Vert^2_{\mathscr H^+(\mathbb R)}}\,.
\end{equation*}
Therefore
\begin{equation}
\label{hat_tf_norm_low_bnd_eqn}
 \Vert \hat T_F^{-1}\Vert=\sup_{\Vert\varphi\Vert_{\mathscr H^+(\mathbb R)}=1}(\varphi,
 \hat T_F^{-1}\varphi)\geq 
 \frac{\Vert\psi_\mu\Vert^2_{\mathcal H}}{\Vert
   x_\mu\Vert^2_{\mathscr H^+(\mathbb R)}}\,.
\end{equation}
Furthermore, according to Eq. (\ref{theta_expression_eqn}) and Eq. (\ref{open_omega_exp_eqn}) we have
\begin{equation}
\label{psi_mu_norm_eqn}
 \Vert\psi_\mu\Vert^2_{\mathcal H}=\Vert U^*\theta
 x_\mu\Vert^2_{\mathcal H}=\Vert\theta x_\mu\Vert^2_{L^2(\mathbb R^+)}=\int_0^\infty
 d\lambda\,\vert\lambda-\mu\vert^{-2}
\end{equation}
and 
\begin{equation*}
 \Vert x_\mu\Vert^2_{\mathscr H^+(\mathbb R)}=\int_{-\infty}^\infty
 d\lambda\,\vert\lambda-\mu\vert^{-2}=\frac{\pi}{\vert\text{Im}\,\mu\vert}\,.
\end{equation*}
Now, changing variables in Eq. (\ref{psi_mu_norm_eqn}) we get that
\begin{equation*}
 \Vert\psi_\mu\Vert^2_{\mathcal H}=\int_{-\text{Re}\,\mu}^\infty
 d\lambda\,\vert\lambda-i\,\text{Im}\,\mu\vert^{-2}
\end{equation*}
and therefore for $\text{Im}\,\mu$ constant we have
$\lim_{\text{Re}\,\mu\to -\infty}\Vert\psi_\mu\Vert^2=0$ and
$\lim_{\text{Re}\,\mu\to\infty}\Vert\psi_\mu\Vert^2
=\Vert x_\mu\Vert^2_{\mathscr H^+(\mathbb R)}$. Hence, taking the limit
$\text{Re}\,\mu\to\infty$ in Eq. (\ref{hat_tf_norm_low_bnd_eqn}) with 
constant $\text{Im}\,\mu$ we obtain that
in fact $\Vert\hat T_F^{-1}\Vert=1$ and, by the equality of the spectrum,
$\Vert\mathbf T_F^{-1}\Vert=1$. Thus, we get that
$\sup\sigma(\mathbf T_F^{-1})=\sup\sigma(\hat T_F^{-1})=1$ and therefore
$\inf\sigma(\mathbf T_F)=\inf\sigma(\hat T_F)=1$. The considerations for the case of the operators
$\mathbf T_B$ and $\hat T_B$ are similar 
with the main difference being in the replacement of $x_\mu$ by a state $x'_\mu\in\mathscr H^-(\mathbb R)$ 
of the form $x'_\mu(\lambda)=(\lambda-\overline\mu)^{-1}$, $\text{Im}\,\mu<0$ and in the replacement of $\psi_\mu$ by the 
state $\psi'_{\mu}:=\Omega_b^*x'_\mu$. One then arrives at the result that 
$\inf\sigma(\mathbf T_B)=\inf\sigma(\hat T_B)=1$.
\par In order to prove the rest of the statements in the theorem we need the following lemma
\begin{lemma}
\label{intertwining_lemma}
The following intertwining relations hold for $t\geq 0$:
\begin{eqnarray}
 \Omega_f\mathbf U(t)&=&T^+_u(t)\Omega_f\\
 \mathbf U(t)\Omega_f^{-1}\vert_{\text{Ran}\, \Omega_f}&=&
 \Omega_f^{-1}T^+_u(t)\vert_{\text{Ran}\, \Omega_f},\label{inv_omega_eqn}\\
 \Omega_f^*T^{+*}_u(t)&=&\mathbf U(-t)\Omega_f^*,\\
 T^{+*}_u(t)(\Omega_f^*)^{-1}\vert_{\text{Ran}\, \Omega_f^*}&=&
 (\Omega_f^*)^{-1}\mathbf U(-t)\vert_{\text{Ran}\, \Omega_f^*}\,.\label{inv_omega_star_eqn}
\label{omega_intertwining_eqn}
\end{eqnarray}
\hfill$\square$
\end{lemma} 
\par{\bf Proof of Lemma \ref{intertwining_lemma}:}
\smallskip
\par Given the mapping $\Omega_f: \mathcal H\mapsto\mathscr H^+(\mathbb R)$, 
consider the basic intertwining relation in Eq. (\ref{basic_intertwining_rel_eqn}). Using this equation we obtain
\begin{multline*}
 (\mathbf U(-t)\Omega_f^*g,f)_{\mathcal H}=(g,\Omega_f\mathbf U(t)f)_{\mathscr H^+(\mathbb R)}
 =(g,T^+_u(t)\Omega_f f)_{H^+(\mathbb R)}
 =(\Omega_f^*T^{+*}_u(t)g,f)_{\mathcal H},\\
 \qquad \forall g\in\mathscr H^+(\mathbb R),\ \forall f\in\mathcal H,\ t\geq 0\\
\end{multline*}
and so
\begin{equation}
\label{t_star_intertwining_eqn}
 \Omega_f^*T^{+*}_u(t)=\mathbf U(-t)\Omega_f^*,\qquad t\geq 0\,.
\end{equation}
By the injective property of the quasi-affine mappings $\Omega_f$
and $\Omega_f^*$, Eq. (\ref{inv_omega_eqn}) and Eq. (\ref{inv_omega_star_eqn}) are direct consequences of 
Eq. (\ref{basic_intertwining_rel_eqn}) and
Eq. (\ref{t_star_intertwining_eqn}).\hfill$\blacksquare$
\par{}
\bigskip
Denote
\begin{equation*}
 T_t(X):=(T^+_u(t))^*XT^+_u(t),\quad t\geq 0,\ X\in\mathcal B(\mathscr
 H^+(\mathbb R))
\end{equation*}
where $\mathcal B(\mathscr H^+(\mathbb R))$ is the space of bounded
operators on $\mathscr H^+(\mathbb R)$.
Using Lemma \ref{intertwining_lemma} we obtain, for any 
$X\in\mathcal B(\mathscr H^+(\mathbb R))$ and $g\in\mathcal H$,
\begin{equation*}
 \Omega_f^*T_t(X)\Omega_f\,g
 =\Omega_f^*(T^+_u(t))^*XT^+_u(t)\Omega_f\,g
 =\mathbf U(-t)\Omega_f^*X\Omega_f\mathbf U(t)\,g\,.
\end{equation*}
Hence,
\begin{equation}
\label{Tt_to_heisenberg_A_eqn}
 (\Omega_f g,T_t(X)\Omega_f g)_{\mathscr H^+(\mathbb R)}=(\mathbf U(t)g,\Omega_f^*X\Omega_f\mathbf U(t)g)_{\mathcal H}\,.
\end{equation}
Denote $g_+:=\Omega_f g$, $g_+(t):=T^+_u(t)g_+$ and
$\tilde g_+(t):=g_+(t)\Vert g_+(t)\Vert_{\mathscr H^+}^{-1}$ and note that, for all $g\in\mathcal H$, 
$\Vert g_+(t_2)\Vert_{\mathscr H^+}\leq\Vert g_+(t_1)\Vert_{\mathscr H^+}$ for $t_2\geq
t_1$ and, for all $g\in\mathcal H$, $\lim_{t\to\infty}\Vert g_+(t)\Vert_{\mathscr H^+}=0$. Then Eq. (\ref{Tt_to_heisenberg_A_eqn})
can be written 
\begin{equation}
\label{normalized_Tt_to_heisen_eqn}
 \Vert g_+(t)\Vert_{\mathscr H^+}^2(\tilde g_+(t),X\tilde
 g_+(t))_{\mathscr H^+}
 =(\mathbf U(t)g,\Omega_f^*X\Omega_f\mathbf U(t)g)_{\mathcal H}\,.
\end{equation}
\par For the operator $X\in\mathcal B(\mathscr H^+(\mathbb R))$ in 
Eq. (\ref{normalized_Tt_to_heisen_eqn}) consider the choice 
$X=\xi_{\mathscr H^+}([1,a))\hat T_F$
(recall that $\xi_{\mathscr H^+}$ is the spectral measure of $\hat T_F$).
For this choice of $X$ Eq. (\ref{normalized_Tt_to_heisen_eqn}) reads
\begin{multline}
\label{hat_TF_estimate_A_eqn}
 \Vert g_+(t)\Vert_{\mathscr H^+}^2(\tilde g_+(t),\xi_{\mathscr H^+}([1,a))
 \hat T_F\xi_{\mathscr H^+}([1,a))\tilde g_+(t))_{\mathscr H^+}=\\
 =(\mathbf U(t)g,\Omega_f^*\hat T_F\xi_{\mathscr
   H^+}([1,a))\Omega_f\mathbf U(t)g)_{\mathcal H}=\\
 =(\mathbf U(t)g,\Omega_f^{-1}\xi_{\mathscr
   H^+}([1,a))\Omega_f\mathbf U(t)g)_{\mathcal H}
 =(\mathbf U(t)g,\xi_{\mathcal H}([1,a))\mathbf U(t)g)_{\mathcal H}\,.\\
\end{multline}
Choose any $g\in\mathcal H$ such that $g=\xi_{\mathcal H}([1,a))g$. If we
also have $\xi_{\mathcal H}([1,a))\mathbf U(t)g=\mathbf U(t)g$ for all $t\geq 0$ then
the right hand side of Eq. (\ref{hat_TF_estimate_A_eqn}) is equal to 
$\Vert g\Vert^2_{\mathcal H}$ for all $t\geq 0$. 
However, $\Vert g_+(t)\Vert_{\mathscr H^+}$ is non-increasing, 
$\lim_{t\to\infty}\Vert g_+(t)\Vert_{\mathscr H^+}=0$ and $\Vert\tilde
g_+(t)\Vert_{\mathscr H^+}=1$, hence there must exist a time $\tau>0$ such that
\begin{equation}
\label{g_decreasing_eqn}
 \Vert g_+(t)\Vert^2_{\mathscr H^+}\bigg(\sup_{\stackrel{t\geq 0}{g\in\mathcal H}}
 \big\vert (\tilde g_+(t),\xi_{\mathscr H^+}([1,a))\hat
 T_F\xi_{\mathscr H^+}([1,a))\tilde g_+(t))_{\mathscr H^+}\big\vert\bigg)
 <\Vert g\Vert^2_{\mathcal H},\quad t>\tau\,.
\end{equation}
The contradiction thus obtained implies that $\xi_{\mathcal
  H}([a,\infty))\mathbf U(t)g\not=\{0\}$ for all $t>\tau$. Furthermore,
since the left hand side of Eq. (\ref{hat_TF_estimate_A_eqn}) vanishes in
the limit as $t$ goes to infinity we must have
$\lim_{t\to\infty}\Vert\xi_{\mathcal H}([1,a))\mathbf U(t)g\Vert_{\mathcal H}=0$. 
\par The proof of the last statement in Theorem \ref{time_observable_thm} concerning the operator $\mathbf T_B$ 
is similar to the proof above for $\mathbf T_F$.\hfill$\blacksquare$
\par\bigskip\
Theorem \ref{time_observable_thm} motivates our point of view of $\mathbf T_F$
as being a time observable for the quantum evolution in the forward
direction since, \emph{for $g\in\mathcal H$, $\mathbf U(t)g$
must ``go up'' on the spectrum of $\mathbf T_F$ as time increases}. We note that
the rate of flow of the evolved state up on the spectrum of $\mathbf T_F$
depends on the choice of the state $g$.
\subsection{The origin of the time observables}
\label{origin_time_observe_subsec}
\par It is clear from the proof of Theorem \ref{time_observable_thm},
and in particular from Eq. (\ref{hat_TF_estimate_A_eqn}) and the
definition of $g_+(t)$, that the existence of the time observables and
the fundamental time asymmetry inherent in the definition of distinct
backward and forward time observables is a direct consequence of the
the fundamental intertwining relations in
Eqns. (\ref{basic_intertwining_rel_eqn}), (\ref{back_basic_intertwine_eqn}). In fact, we can do better and
show that further analysis of this basic equation provides a more
clear understanding of the origin of the time observables. We take up this
task in this subsection (the discussion here partly follows Section
(IV) of reference \cite{St2}). 
\par Let $\mathcal S$ be the \emph{Schwartz class} of rapidly
decreasing functions in $C^\infty(\mathbb R)$ and let $\mathcal S'$ be
the space of \emph{tempered distributions} on $\mathcal S$. For any
fixed $p\in(0,\infty)$ let $\mathscr H^p(\mathbb C\backslash\mathbb
R)$ be the space of analytic functions on $\mathbb C\backslash\mathbb
R$ for which
\begin{equation*}
 \Vert f\Vert=\sup_{y\not= 0}\left(\int_{\mathbb R}\vert
   f(x+iy)\vert^p\right)^{1/p}<\infty\,.
\end{equation*}
It can be shown \cite{CR} that every function $F\in\mathscr H^p(\mathbb C\backslash\mathbb R)$ 
is associated with a unique tempered distribution $\ell_F\in\mathcal S'$ defined by
\begin{equation}
\label{F_distribution_eqn}
 \ell_F(\psi)=\lim_{y\to 0^+}\int_{\mathbb
   R}\{F(x+iy)-F(x-iy)\}\psi(x)dx,\qquad \psi\in\mathcal S\,.
\end{equation}
We denote the set of all such distributions by $H^p(\mathbb
R)$. Conversly, to any distribution $\ell\in H^p(\mathbb R)$ with
$p\in(0,\infty)$ we can associate a unique function $F_\ell\in\mathscr
H^p(\mathbb C\backslash\mathbb R)$ such that 
\begin{equation*}
 \ell_{F_\ell}(\psi)=\lim_{y\to 0^+}\int_{\mathbb
   R}\{F_\ell(x+iy)-F_\ell(x-iy)\}\psi(x)dx=\ell(\psi)
 ,\quad \psi\in\mathcal S\,.
\end{equation*}
The function $F_\ell\in\mathscr H^p(\mathbb C\backslash\mathbb R)$ is then given by \cite{CR}
\begin{equation}
\label{dist_to_func_eqn}
 F_\ell(z)=\frac{1}{2\pi i}\ell((\cdot-z)^{-1}),\qquad z\in\mathbb
 C\backslash\mathbb R\,.
\end{equation}
Now, for $p\in(1,\infty)$ we have the further identification of the
space of distributions $H^p(\mathbb R)$ with the function space
$L^p(\mathbb R)$ in the sense that any function $f\in L^p(\mathbb R)$
defines a tempered distribution on $\mathcal S$ via
\begin{equation}
\label{Lp_dist_eqn}
 \ell_f(\psi)=\int_{\mathbb R}f(x)\psi(x)\,dx,\qquad \psi\in\mathcal S
\end{equation}
and Eq. (\ref{dist_to_func_eqn}) associates with $f$ a unique analytic function 
$F_{\ell_f}\in\mathscr H^p(\mathbb C\backslash\mathbb R)$, i.e.,
\begin{equation}
\label{Lp_dist_to_func_eqn}
 F_{\ell_f}(z)=\frac{1}{2\pi 1}\ell_f((\cdot-z)^{-1})
 =\frac{1}{2\pi i}\int_{\mathbb R}\frac{f(x)}{x-z}\,dx\,.
\end{equation}
Using Eq. (\ref{F_distribution_eqn}) we can then recover the
distribution $\ell_f$ from the function $F_{\ell_f}$, i.e., we have $\ell_{F_{\ell_f}}=\ell_f$.
For our purpose we need also the following proposition \cite{CR}:
\begin{proposition}
\label{interval_omission_prop}
A distribution $\ell\in H^p(\mathbb R)$ has support which omits an
open interval $\Delta\in\mathbb R$ iff the corresponding function
$F_\ell\in\mathscr H^p(\mathbb C\backslash\mathbb R)$ given by
Eq. (\ref{dist_to_func_eqn}) has an anlytic continuation across the
interval $\Delta$.
\end{proposition}
We now restrict the discussion above to the case $p=2$ and consider
the embedding $I:L^2(\mathbb R^+)\mapsto L^2(\mathbb R)$ in Eq. (\ref{embedding_eqn}). For
any $f\in L^2(\mathbb R^+)$ we have $If\in L^2(\mathbb
R)$. Using the identification of $L^2(\mathbb R)$ with the space
$H^2(\mathbb R)$ as above we associate with $If$, through
Eq. (\ref{Lp_dist_to_func_eqn}), the function $F_{\ell_{If}}\in\mathscr H^p(\mathbb C\backslash\mathbb R)$.
Clearly the distribution defined by $If$ omits the interval $\mathbb
R^-$ and therefore, by Proposition \ref{interval_omission_prop},
$F_{\ell_{If}}$ is analytic across the
neagtive real axis i.e, $F_{\ell_{If}}\in\mathscr H^2(\mathbb
C\backslash\mathbb R^+)\subset\mathscr H^2(\mathbb
C\backslash\mathbb R)$ where $\mathscr H^2(\mathbb
C\backslash\mathbb R^+)$ denotes the subspace of $\mathscr H^2(\mathbb
C\backslash\mathbb R)$ containing functions analytic across $\mathbb
R^-$. We shall use the notation $F_f\equiv F_{\ell_{If}}$.
\par Note that Eq. (\ref{F_distribution_eqn}) and the uniqueness of
the functional $\ell_f$ in Eq. (\ref{Lp_dist_eqn}) allows us to associate
with each function $F\in\mathscr H^2(\mathbb C\backslash\mathbb R^+)$
the corresponding function $f\in L^2(\mathbb R^+)$. Thus we have the
following lemma
\begin{lemma}
\label{A_prime_definiton_eqn}
There exists a bijective map $A':L^2(\mathbb R^+)\mapsto\mathscr
H^2(\mathbb C\backslash\mathbb R^+)$ such that $A'f=F_f$ with $f\in
L^2(\mathbb R^+)$, $F_f\in\mathscr H^2(\mathbb C\backslash\mathbb
R^+)$, $F_f$ given by 
\begin{equation*}
 F_f(z)=\frac{1}{2\pi i}\int_{\mathbb
   R^+}\frac{f(x)}{x-z}\,dx,\qquad f\in L^2(\mathbb R^+)\,,
\end{equation*}
and, given $F_f$ we have
\begin{equation}
\label{reconstruct_f_eqn}
 f=A'^{-1}F_f=I^{-1}[F^+_f-F^-_f]
\end{equation}
where 
\begin{equation}
\label{F_f_boundary_value_eqn}
 F^+_f(\sigma)=\lim_{\epsilon\to 0^+}F_f(\sigma+i\epsilon),\qquad
 F^-_f(\sigma)=\lim_{\epsilon\to 0^+}F_f(\sigma-i\epsilon)\,\qquad \sigma\in\mathbb R\,.
\end{equation}
and we note that the boundary value functions $F_f^+$ and $F_f^-$ exist a.e.
since the restriction of $F_f$ to $\mathbb C^+$ is an element of
$\mathscr H^2(\mathbb C^+)$ and the restriction of $F_f$ to $\mathbb
C^-$ is an element of $\mathscr H^2(\mathbb C^-)$.
\hfill$\square$
\end{lemma}
\par{\bf Proof:}
\smallskip
\par In view of the discussion above we have only to find an explicit
form for the transformation $A$. This is obtained through the use of
Eq. (\ref{Lp_dist_to_func_eqn}) with the result
\begin{equation*}
 F_f(z)\equiv F_{\ell_{If}}(z)=\frac{1}{2\pi i}\int_{\mathbb
   R}\frac{If(x)}{x-z}\,dx
 =\frac{1}{2\pi i}\int_{\mathbb
   R^+}\frac{f(x)}{x-z}\,dx,\qquad f\in L^2(\mathbb R^+)\,.
\end{equation*}
In addition Eq. (\ref{reconstruct_f_eqn}) is a direct result of Eq. (\ref{F_distribution_eqn}) and Eq. (\ref{Lp_dist_eqn}).
\hfill$\blacksquare$
\par\bigskip\
Consider the unitary map $U: \mathcal H\mapsto L^2(\mathbb
R^+)$ mapping $\mathcal H$ onto its energy
representation on the spectrum of $\mathbf H$. Combining the mappings
$U$ and $A'$ we get a bijective map $A:\mathcal H\mapsto
\mathscr H^2(\mathbb C\backslash\mathbb R^+)$ with $A=:A'U$. For
an element $\psi\in\mathcal H$ denote
$\psi_A=A\psi=A'U\psi$. Choosing an element $\psi\in\mathcal H$ as an initial state
and letting it evolve under the Schr\"odinger evolution $\mathbf U(t)$
we get an induced evolution in $\mathscr H^2(\mathbb C\backslash\mathbb R^+)$
\begin{equation*}
 \psi_A(t)=A\psi_t=A'U\psi_t=A'U\mathbf U(t)\psi\,.
\end{equation*}
We would like to characterize this induced evolution. Denote by
$\psi_A^+(t)$ the restriction of $\psi_A(t)$ to $\mathbb C^+$ and note
that $\psi_A^+(t)\in\mathscr H^2(\mathbb C^+)$. Similarly, if
$\psi_A^-(t)$ denotes the restriction of $\psi_A(t)$ to $\mathbb C^-$
then $\psi_A^-(t)\in\mathscr H^2(\mathbb C^-)$. For each time $t$ we
have
\begin{equation*}
 \Vert\psi_A^+(t)\Vert_{\mathscr H^2(\mathbb C^+)}<\infty,\quad
 \Vert\psi_A^-(t)\Vert_{\mathscr H^2(\mathbb C^-)}<\infty\,.
\end{equation*}
Recall that for an element $f\in L^2(\mathbb R^+)$ 
the boundary value functions $F^+_f$ and $F^-_f$ (see Eq. (\ref{F_f_boundary_value_eqn}) above) 
belong, repectively, to $\mathscr H^+(\mathbb R)$ and
$\mathscr H^-(\mathbb R)$. Considering the mappings $\theta^*: L^2(\mathbb R^+)\mapsto\mathscr
H^+(\mathbb R)$ and $\overline\theta^*: L^2(\mathbb R^+)\mapsto\mathscr
H^-(\mathbb R)$ for any function $f\in L^2(\mathbb R^+)$ we have
\begin{equation}
\label{L2R_decomp_eqn}
 If=P_+If+P_-If=\theta^*f+\overline\theta^*f\,.
\end{equation}
Since $L^2(\mathbb
R)=\mathscr H^+(\mathbb R)\oplus\mathscr H^-(\mathbb R)$ the
sum in Eq. (\ref{L2R_decomp_eqn}) is unique. However, from Eq. (\ref{F_f_boundary_value_eqn})
we obtain
\begin{equation*} 
 If=IA'^{-1}F_f=F^+_f-F^-_f\,.
\end{equation*}
hence $F^+_f=\theta^*f$ and $F^-_f=-\overline\theta^*f$. Denote by
$F^+_{\psi_A(t)}$ the boundary value of $\psi_A^+(t)$ on $\mathbb R$
and by $F^-_{\psi_A(t)}$ the boundary value of $\psi_A^-(t)$ on
$\mathbb R$. Then $F^+_{\psi_A(t)}\in\mathscr H^+(\mathbb R)$ and  
$F^-_{\psi_A(t)}\in\mathscr H^-(\mathbb R)$ and we have, for any
$t\in\mathbb R$
\begin{equation*}
 F^+_{\psi_A(t)}=\theta^*U\psi_t=\Omega_f\mathbf U(t)\psi
\end{equation*}
and 
\begin{equation*}
 F^-_{\psi_A(t)}=-\overline\theta^*U\psi_t=-\Omega_b\mathbf U(t)\psi\,.
\end{equation*}
Using Eq. (\ref{basic_intertwining_rel_eqn}) and the isomorphism of
$\mathscr H^2(\mathbb C^+)$ and $\mathscr H^+(\mathbb R)$ we get
\begin{equation}
\label{psi_A_plus_norm_eqn}
 \Vert\psi_A^+(t)\Vert_{\mathscr H^2(\mathbb C^+)}
 =\Vert F^+_{\psi_A(t)}\Vert_{\mathscr H^+(\mathbb R)}
 =\Vert\Omega_f\mathbf U(t)\psi\Vert_{\mathscr H^+(\mathbb R)}
 =\Vert T^+_u(t)\Omega_f\psi\Vert_{\mathscr H^+(\mathbb R)},\quad t\geq 0\,.
\end{equation}
We conclude that, for $t\geq 0$, $\Vert\psi_A^+(t)\Vert_{\mathscr
  H^2(\mathbb C^+)}$ is monotonically decreasing and,
furthermore, $\lim_{t\to\infty}\Vert\psi_A^+(t)\Vert_{\mathscr H^2(\mathbb C^+)}=0$.
Because of the fact that $\Vert\psi_A^+(t)\Vert_{\mathscr H^2(\mathbb
  C^+)}+\Vert\psi_A^-(t)\Vert_{\mathscr H^2(\mathbb
  C^-)}=\Vert\psi\Vert_{\mathcal H}$ we obtain that
$\Vert\psi_A^-(t)\Vert_{\mathscr H^2(\mathbb C^-)}$ is monotonically
increasing for $t\geq 0$ and we have
$\lim_{t\to\infty}\Vert\psi_A^-(t)\Vert_{\mathscr H^2(\mathbb
  C^-)}=\Vert\psi\Vert_{\mathcal H}$.
Using Eq. (\ref{back_basic_intertwine_eqn}) and the isomorphism of
$\mathscr H^2(\mathbb C^-)$ and $\mathscr H^-(\mathbb R)$ we obtain
\begin{equation}
\label{psi_A_minus_norm_eqn}
 \Vert\psi_A^-(t)\Vert_{\mathscr H^2(\mathbb C^-)}
 =\Vert F^-_{\psi_A(t)}\Vert_{\mathscr H^-(\mathbb R)}
 =\Vert\Omega_b\mathbf U(t)\psi\Vert_{\mathscr H^-(\mathbb R)}
 =\Vert T^-_u(t)\Omega_b\psi\Vert_{\mathscr H^-(\mathbb R)},\quad t\leq 0\,.
\end{equation}
and we conclude that, for $t\leq 0$, $\Vert\psi_A^-(t)\Vert_{\mathscr
  H^2(\mathbb C^-)}$ is monotonically decreasing and we have
$\lim_{t\to -\infty}\Vert\psi_A^-(t)\Vert_{\mathscr H^2(\mathbb C^-)}=0$.
Moreover, $\Vert\psi_A^+(t)\Vert_{\mathscr H^2(\mathbb C^+)}$ is monotonically
increasing for $t\leq 0$ and 
$\lim_{t\to -\infty}\Vert\psi_A^+(t)\Vert_{\mathscr H^2(\mathbb
  C^+)}=\Vert\psi\Vert_{\mathcal H}$.
Summarising, we have proved the following 
\begin{proposition}
\label{norm_flow_prop}
There exists a bijective map $A:\mathcal H\mapsto\mathscr H^2(\mathbb C\backslash\mathbb R^+)$
with $A=A'U$, $U:\mathcal H\mapsto L^2(\mathbb R^+)$ is the mapping of $\mathcal H$ onto its
energy representation on the spectrum of $\mathbf H$ and 
$A':L^2(\mathbb R^+)\mapsto\mathscr H^2(\mathbb C\backslash\mathbb R^+)$
defined in Lemma \ref{A_prime_definiton_eqn}.
For $\psi\in\mathcal H$ denote $\psi_A=A\psi$
and $\psi_A(t)=A\mathbf U(t)\psi$. Denote 
$\psi_A^+(t)$ the restriction of $\psi_A(t)$ to $\mathbb C^+$
and $\psi_A^-(t)$ the restriction of $\psi_A(t)$ to $\mathbb C^-$.
Then we have 
\begin{equation*}
 \Vert\psi_A^+(t_1)\Vert_{\mathscr H^2(\mathbb C^+)}\geq 
 \Vert\psi_A^+(t_2)\Vert_{\mathscr H^2(\mathbb C^+)},\quad 0\leq t_1<t_2\,,
\end{equation*} 
\begin{equation*}
 \Vert\psi_A^-(t_1)\Vert_{\mathscr H^2(\mathbb C^-)}\leq 
 \Vert\psi_A^-(t_2)\Vert_{\mathscr H^2(\mathbb C^-)},\quad 0\leq
 t_1<t_2
\end{equation*} 
and 
\begin{equation*}
 \lim_{t\to\infty}\Vert\psi_A^+(t)\Vert_{\mathscr H^2(\mathbb
   C^+)}=0,\quad
 \lim_{t\to\infty}\Vert\psi_A^-(t)\Vert_{\mathscr H^2(\mathbb
   C^-)}=\Vert\psi\Vert_{\mathcal H}\,.
\end{equation*}
In addition
\begin{equation*}
 \Vert\psi_A^+(t_1)\Vert_{\mathscr H^2(\mathbb C^+)}\leq 
 \Vert\psi_A^+(t_2)\Vert_{\mathscr H^2(\mathbb C^+)},\quad t_2<t_1\leq 0\,,
\end{equation*} 
\begin{equation*}
 \Vert\psi_A^-(t_1)\Vert_{\mathscr H^2(\mathbb C^-)}\geq 
 \Vert\psi_A^-(t_2)\Vert_{\mathscr H^2(\mathbb C^-)},\quad t_2<t_1\leq 0
\end{equation*} 
and 
\begin{equation*}
 \lim_{t\to -\infty}\Vert\psi_A^-(t)\Vert_{\mathscr H^2(\mathbb
   C^-)}=0,\quad
 \lim_{t\to -\infty}\Vert\psi_A^+(t)\Vert_{\mathscr H^2(\mathbb
   C^+)}=\Vert\psi\Vert_{\mathcal H}\,.
\end{equation*}
\hfill$\square$
\end{proposition}
It is the flow of norm from the upper half-plane Hardy space to the
lower half-plane Hardy space induced by the Schr\"odinger evolution for positive times that gives rise to the time
observable for forward evolution. In fact, as is evident from the proof of Proposition
\ref{norm_flow_prop}, this flow of norm provides the basic
intertwining relation Eq. (\ref{basic_intertwining_rel_eqn}) which in
turn stands at the heart of the proof of Theorem \ref{time_observable_thm}.
In a similar manner, for negative times the Schr\"odinger evolution induces 
the flow of norm from the lower half-plane Hardy space to
the upper half-plane Hardy space which gives rise to the time observable for backward evolution.
\section{Time observables for quantum stochastic processes}
\label{t_observe_stochastic_process_sec}
\subsection{Mapping of creation, annihilation and conservation processes}
\label{creation_annihilation_conservation_map_subsec}
\par As mentioned in Section \ref{introduction} there exists an inherent time asymmetry built into the semigroup 
decomposition formalism in the form of two distinct semigroup evolutions appearing in Eq. (\ref{basic_intertwining_rel_eqn})
and Eq. (\ref{back_basic_intertwine_eqn}) and 
corresponding respectively to future directed evolution and to past directed evolution and in the existence of distinct
forward and backward time observables $\mathbf T_F$ and $\mathbf T_B$. Considering for the moment forward time evolution
(the treatment of backward evolution parallels the developments below) one may ask, in light of the discussion in Section
\ref{the_time_observables_sec}, whether the use of $\mathbf T_F$ can be extended in such a way that it may serve a universal role 
as a forward time observable for more general quantum processes. Following this line of thought we will consider in this
section the role of $\mathbf T_F$ as a time observable for quantum stochastic processes. We shall work in the setting of
quantum stochastic differential equations defined in the framework of the Hudson-Parthasarathy (HP) quantum stochastic calculus.
The terminology and notation below closely follows that of \cite{Par2}. 
\par A simple answer to the question whether $\mathbf T_F$ can be applied as a time observable for quantum 
stochastic processes is: yes. This stems from the fact that on the abstract level a general $\mathbb R^+$-valued observable, i.e., a self-adjoint
operator with spectral projection valued measure $\xi$ defined on the Borel $\sigma$-algebra $\mathfrak B^+$, can be used as a
time observable with respect to which one may define $\xi$-martingales
and basic regular adapted processes which are then utilized for the definition of stochastic integration and the construction of 
quantum stochastic differential equations \cite{HuP}. This abstract requirement is, however, not informative in the sense that it gives no characterization
of the nature of the self-adjoint operator playing the role of a time observable. Therefore, a more concrete question is whether one may find the analogue of the
fundamental intertwining relation in Eq. (\ref{basic_intertwining_rel_eqn}) (and Eq. (\ref{back_basic_intertwine_eqn}) for the backward case). In other words one may ask 
whether it is possible to find a map associated with $\Omega_f$ that intertwines a (quantum) stochastic process, defined with
respect to the physical Hilbert space $\mathcal H$ and the time observable $\mathbf T_F$, with a (quantum)
stochastic process defined with respect to the Hardy space $\mathscr H^+(\mathbb R)$ and the observable $\hat T_F$. We
address this question in the present section.
\par As in Section \ref{the_time_observables_sec} above let $\xi_{\mathcal H}:\mathfrak B^+\mapsto
\mathcal P(\mathcal H)$ be the spectral measure of $\mathbf T_F$ and $\xi_{\mathscr H^+}:\mathfrak B^+
\mapsto\mathcal P(\mathscr H^+(\mathbb R))$ be the spectral measure of $\hat T_F$. The first step in the construction of   
fundamental adapted processes which respect to which stochastic integration can be defined is the definition of 
$\xi_{\mathcal H}$-martingales and of $\xi_{\mathscr H^+}$-martingales. We first recall the definition of $\xi$-martingales. Let
$\mathcal K$ be a complex separable Hilbert space and let $\xi: \mathfrak B^+\mapsto\mathcal P(\mathcal K)$
be a fixed $\mathbb R^+$-valued observable. For $0\leq s<t$ we define
\begin{equation*}
 \mathcal K_{t]}:=\xi([0,t])\mathcal K,\quad 
 \mathcal K_{[s,t]}:=\xi([s,t])\mathcal K,\quad
 \mathcal K_{[t}:=\xi([t,\infty))\mathcal K\,.
\end{equation*}
Then a $\xi$-martingale on $\mathcal K$ is defined as follows
\begin{definition}
\label{martingale_def}
Let $\xi$ be an $\mathbb R^+$-valued observable on $\mathcal K$. Let $m:\mathbb R^+\mapsto\mathcal K$ be a map and 
for $t\in\mathbb R^+$ denote $m(t)\equiv m_t$. If the map $m$ satisfies:
\begin{enumerate}
\item $m_t\in\mathcal K_{t]}$, $\forall t\geq 0$,
\item $\xi([0,s])m_t=m_s$, $s<t$,
\end{enumerate}
then $m$ is called a $\xi$-martingale.\hfill$\square$
\end{definition}
In the case of the Hilbert spaces $\mathcal H$ and $\mathscr H^+(\mathbb R)$ and the time observables 
$\mathbf T_F$ and $\hat T_F$ we shall use the notation $\mathcal H_{t]}=\xi_{\mathcal H}([1,t+1])\mathcal H$,
$\mathcal H_{[s,t]}=\xi_{\mathcal H}([s+1,t+1])\mathcal H$, 
$\mathcal H_{[t}=\xi_{\mathcal H}([t+1,\infty))\mathcal H$ and 
$\mathscr H^+_{t]}=\xi_{\mathscr H^+}([1,t+1])\mathscr H^+(\mathbb R)$,
$\mathscr H^+_{[s,t]}=\xi_{\mathscr H^+}([s+1,t+1])\mathscr H^+(\mathbb R)$, 
$\mathscr H^+_{[t}=\xi_{\mathscr H^+}([t+1,\infty))\mathscr H^+(\mathbb R)$.
Then, a  $\xi_{\mathcal H}$-martingale is defined as in Definition \ref{martingale_def} with 
conditions (.1)-(.2) adjusted in the form
\begin{itemize}
\item[$(1.)_{\mathcal H}$] $m_t\in\mathcal H_{t]}$, $\forall t\geq 0$
\item[$(2.)_{\mathcal H}$] $\xi_{\mathcal H}([1,s+1])m_t=m_s$, $s<t$,
\end{itemize}
and a $\xi_{\mathscr H^+}$-martingale is defined as in Definition \ref{martingale_def} with conditions (.1)-(.2) adjusted in the form
\begin{itemize}
\item[$(1.)_{\mathscr H}$] $m_t\in\mathscr H^+_{t]}$, $\forall t\geq 0$
\item[$(2.)_{\mathscr H}$] $\xi_{\mathscr H^+}([1,s+1])m_t=m_s$, $s<t$,
\end{itemize}
Having defined $\xi_{\mathcal H}$-martingales and $\xi_{\mathscr H^+}$-martingales we have the following lemma
concerning the mappings of martingales
\begin{lemma}
\label{martingale_mapping_lemma}
Let $m$ be a $\xi_{\mathcal H}$-martingale. Then the map $\hat m:\mathbb R^+\mapsto\mathscr H^+(\mathbb R)$
defined by $\hat m_t\equiv \hat m(t):=\Omega_f\,m_t$, $t\geq 0$ is a $\xi_{\mathscr H^+}$-martingale.\hfill$\square$
\end{lemma}
\par{\bf Proof of lemma \ref{martingale_mapping_lemma}:}
\smallskip
\par This lemma is a result of Eq. (\ref{spectral_family_transform_A_eqn}) and 
Eq. (\ref{spectral_family_transform_B_eqn})  in Proposition \ref{time_operators_prop} and the following simple calculations
\begin{equation}
\label{phys_to_hardy_martingale_A_eqn}
 \Omega_f\,\mathcal H_{t]}=\Omega_f\xi_{\mathcal H}([1,t+1])\mathcal H
 =\Omega_f\xi_{\mathcal H}([1,t+1])\Omega_f^{-1}\Omega_f\mathcal H
 =\xi_{\mathscr H^+}([1,t+1])\Omega_f\mathcal H\subset \mathscr H^+_{t]}\,.
\end{equation}
and
\begin{multline}
\label{phys_to_hardy_martingale_B_eqn}
 \xi_{\mathscr H^+}([1,s+1])\hat m_t=(\Omega_f^*)^{-1}\xi_{\mathcal H}([1,s+1])\Omega_f^*\Omega_f m_t
 =(\Omega_f^*)^{-1}\xi_{\mathcal H}([1,s+1])\mathbf T_F^{-1} m_t=\\
 =(\Omega_f^*)^{-1}\mathbf T_F^{-1}\xi_{\mathcal H}([1,s+1]) m_t=\Omega_f m_s=\hat m_s\,.
\end{multline}
Eq. (\ref{phys_to_hardy_martingale_A_eqn}) shows that if $m_t\in\mathcal H_{t]}$ then 
$\hat m_t\in\mathscr H^+_{t]}$ in agreement with condition (1) in Definition \ref{martingale_def}. 
Eq. (\ref{phys_to_hardy_martingale_B_eqn}) corresponds to condition (2) in Definition \ref{martingale_def}.
Hence $\hat m$ is a $\xi_{\mathscr H^+}$-martingale.\hfill$\square$.
\par\bigskip\
\par We now use the mapping of martingales given in Lemma \ref{martingale_mapping_lemma} to map elementary adapted
stochastic processes. We need first the following lemma
\begin{lemma}
\label{K_to_hat_K_lemma}
Let $\mathbf K\in\mathcal B(\mathcal H)$ satisfy $[\mathbf K,\xi_{\mathcal H}(E)]=0$, $\forall E\in\mathfrak B^+$.
Define
\begin{equation}
\label{hat_K_definition_eqn}
 \hat K:=\overline{(\Omega_f^*)^{-1}\mathbf K\Omega_f^*}\,.
\end{equation}
Then $\hat K\in\mathcal B(\mathscr H^+(\mathbb R))$ and $[\hat K,\xi_{\mathscr H^+}(E)]=0$, $\forall E\in\mathfrak B^+$.
\hfill$\square$
\end{lemma}
\par{\bf Proof of Lemma \ref{K_to_hat_K_lemma}:}
\smallskip
\par Denote $\vert\Omega_f\vert=(\Omega_f^*\Omega_f)^{1/2}=(\mathbf T_F^{-1})^{1/2}$. Consider the map
$\vert\Omega_f\vert^{-1}\Omega_f^*$ and note that 
\begin{equation}
\label{X_star_eqn}
 (\vert\Omega_f\vert^{-1}\Omega_f^*)^*=\Omega_f\vert\Omega_f\vert^{-1}
 =(\Omega_f^*)^{-1}\Omega_f^*\Omega_f\vert\Omega_f\vert^{-1}=(\Omega_f^*)^{-1}\vert\Omega_f\vert\,.
\end{equation}
Note further that the map $\vert\Omega_f\vert^{-1}\Omega_f^*$ is well defined on the dense set 
$\text{Ran}\,\Omega_f\subset\mathscr H^+(\mathbb R)$. Using Eq. (\ref{X_star_eqn}) it is clear that this map
be extended to a unitary map $X$ from $\mathscr H^+(\mathbb R)$ to $\mathcal H$ (see \cite{SzNF}). The 
adjoint $X^*$ is then an extension of the right hand side of Eq. (\ref{X_star_eqn}).
Using the unitary extension $X$ we define an operator $K':=X^*\mathbf K X$. Obviously we have $K'\in\mathcal B(\mathscr H^+(\mathbb R))$. 
By assumption $\mathbf K$ commutes with the spectral measure $\xi_{\mathcal H}$ and hence with $\vert\Omega_f\vert$.
Thus, for any $f\in\text{Ran}\,\Omega_f$ we obtain with the help of Eq. (\ref{X_star_eqn})
\begin{equation*}
 K'f=X^*\mathbf K Xf=\Omega_f \vert\Omega_f\vert^{-1}\mathbf K\vert\Omega_f\vert^{-1}\Omega_f^* f
 =(\Omega_f^*)^{-1}\mathbf K\Omega_f^* f\,.
\end{equation*}
Hence we get that $\hat K=\overline{(\Omega_f^*)^{-1}\mathbf K\Omega_f^*}=K'\in\mathcal B(\mathscr H^+(\mathbb R))$. Moreover, using
 Eq.(\ref{spectral_family_transform_B_eqn}) we have
\begin{multline}
\label{K_measure_intertwine_eqn}
 \xi_{\mathscr H^+}(E)(\Omega_f^*)^{-1}\mathbf K\Omega_f^* f
 =(\Omega_f^*)^{-1}\xi_{\mathcal H}(E)\Omega_f^*(\Omega_f^*)^{-1}\mathbf K\Omega_f^*=(\Omega_f^*)^{-1}\xi_{\mathcal H}(E)\mathbf K\Omega_f^*=\\
 =(\Omega_f^*)^{-1}\mathbf K\xi_{\mathcal H}(E)\Omega_f^*=(\Omega_f^*)^{-1}\mathbf K\Omega_f^*\xi_{\mathscr H^+}(E)\,.
\end{multline} 
The commutation relation $[\hat K,\xi_{\mathscr H^+}(E)]=0$ is then obtained by taking the closure of the operator $(\Omega_f^*)^{-1}\mathbf K\Omega_f^*$
in Eq. (\ref{K_measure_intertwine_eqn}).\hfill$\blacksquare$
\par\bigskip\
\par Let $\mathcal H_0$ be a complex separable Hilbert space and let $\Gamma_s(\mathcal H)$
be the symmetric (Bosonic) Fock space over $\mathcal H$ and $\Gamma_s(\mathscr H^+(\mathbb R))$ be the
symmetric Fock space over $\mathscr H^+(\mathbb R)$ \cite{BR,Par2}. Denote
\begin{equation}
\label{tilde_H_tilde_H_plus_eqn}
 \tilde{\mathcal H}=\mathcal H_0\otimes\Gamma_s(\mathcal H),\qquad
 \tilde{\mathscr H}^+=\mathcal H_0\otimes\Gamma_s(\mathscr H^+(\mathbb R))\,.
\end{equation}
Below we consider in $\tilde{\mathcal H}$ \emph{regular adapted processes} with respect to
the triplet $(\xi_{\mathcal H},\mathcal H_0,\mathcal H)$ and in $\tilde{\mathscr H}^+$ regular
adapted processes with respect to the triplet $(\xi_{\mathscr H^+},\mathcal H_0,\mathscr H^+(\mathbb R))$ \cite{HuP,Par2}.
Denote by $a(u)$ the \emph{annihilation operator} and by $a^\dagger(u)$ the \emph{creation operator} in 
$\Gamma_s(\mathcal H)$ associated with $u\in\mathcal H$ and, for $\mathbf K\in\mathcal B(\mathcal H)$,  
denote by $\lambda(\mathbf K)$ the \emph{conservation operator} associated with $\mathbf K$ \cite{BR,Par2}. 
The annihilation, creation and conservation operators in $\Gamma_s(\mathscr H^+(\mathbb R))$ are 
denoted respectively by $\hat a(u)$, $\hat a^\dagger(u)$ and $\hat\lambda(\hat K)$ where $u\in\mathscr H^+(\mathbb R)$ and
$\hat K\in\mathcal B(\mathscr H^+(\mathbb R))$. For $u$ in $\mathcal H$ we denote by $e(u)$
the \emph{exponential vector} in $\Gamma_s(\mathcal H)$ associated with $u$ and by 
$\mathcal E(\mathcal H)$ the linear manifold generated by $\{e(u)\mid u\in\mathcal H\}$. 
The analogous objects in $\Gamma_s(\mathscr H^+(\mathbb R))$ are denoted $\hat e(u)$ and 
$\mathcal E(\mathscr H^+(\mathbb R))$. Note that $\{e(u)\mid u\in\mathcal H\}$ is
\emph{total} in $\Gamma_s(\mathcal H)$
and $\{e(u)\mid u\in\mathscr H^+(\mathbb R)\}$ is total in $\Gamma_s(\mathscr H^+(\mathbb R))$.
\par Let $\mathscr W$ be the algebra of operators on $\Gamma_s(\mathcal H)$ generated by
$\{a(u),a^\dagger(u),\lambda(\mathbf K), \mathbf 1_{\Gamma_s(\mathcal H)}\linebreak[2]\\ \mid u\in\mathcal H
, \mathbf K\in\mathcal B(\mathcal H), [\mathbf K,\mathbf T_F]=0\}$.
Let $\hat{\mathscr W}$ be the corrsponding algebra on $\Gamma_s(\mathscr H^+(\mathbb R))$ generated by 
$\{\hat a(u),\hat a^\dagger(u),\hat\lambda(\hat K),\mathbf 1_{\Gamma_s(\mathscr H^+)}
\mid u\in\mathscr H^+(\mathbb R),\hat K\in\mathcal B(\mathscr H^+(\mathbb R)), [\hat K,\hat T_F]=0\}$.
Then the map $\Omega_f:\mathcal H\mapsto\mathscr H^+(\mathbb R)$ induces a transformation
$\Gamma(\Omega_f):\mathscr W\mapsto\hat{\mathscr W}$ via the definition
\begin{definition}[$\Gamma(\Omega_f)$, mapping of the algebra]
\label{mapping_algebra_def}
Let $\mathbf K$ be an operator in $\mathcal B(\mathcal H)$. Define the operator $\hat K$ according to 
Eq. (\ref{hat_K_definition_eqn}) i.e., $\hat K=\overline{(\Omega_f^*)^{-1}\mathbf K\Omega_f^*}$ and assume that $\mathbf K$ is
such that $\hat K\in\mathcal B(\mathscr H^+(\mathbb R))$. 
Given the quasi-affine map $\Omega_f:\mathcal H\mapsto\mathscr H^+(\mathbb R)$ we define the map
$\Gamma(\Omega_f):\mathscr W\mapsto\hat{\mathscr W}$ through the following relations
\begin{enumerate}
\item $\Gamma(\Omega_f)\mathbf 1_{\Gamma_s(\mathcal H)}=\mathbf 1_{\Gamma_s(\mathscr H^+)}\Gamma(\Omega_f)$,
\item $\Gamma(\Omega_f)a(u)=\hat a(\Omega_f u)\Gamma(\Omega_f)$, $\forall u\in\mathcal H$,
\item $\Gamma(\Omega_f)a^\dagger(u)=\hat a^\dagger(\Omega_fu)\Gamma(\Omega_f)$, $\forall u\in\mathcal H$,
\item $\Gamma(\Omega_f)\lambda(\mathbf K)=\hat\lambda(\hat K)\Gamma(\Omega_f)$.
\end{enumerate}
\hfill$\square$
\end{definition}
Applying the transformation $\Gamma(\Omega_f)$ to an element $X\in\mathscr W$ we obtain an element 
$\hat X\in\hat{\mathscr W}$. We complete the definition of $\Gamma(\Omega_f)$ by determining its action on states defined
on $\mathscr W$ and $\hat{\mathscr W}$: 
\begin{definition}[$\Gamma(\Omega_f)$, mapping of the state]
Let $\Phi_{\mathcal H}$ be the vacuum vector in
$\Gamma_s(\mathcal H)$ and let $\Phi_{\mathscr H^+}$ be the vacuum vector in 
$\Gamma_s(\mathscr H^+(\mathbb R))$. We define the action of $\Gamma(\Omega_f)$ on $\Phi_{\mathcal H}$ by
\begin{equation*}
 \Gamma(\Omega_f)\Phi_{\mathcal H}=\Phi_{\mathscr H^+}\,.
\end{equation*}
\hfill$\square$
\end{definition}
As an example consider the application of $\Gamma(\Omega_f)$ to the exponential vector 
$e(u)\in\Gamma_s(\mathcal H)$. For this vector we have a representation in the form
$e(u)=\exp[a^\dagger(u)]\Phi_{\mathcal H}$. Hence
\begin{equation*}
 \Gamma(\Omega_f)e(u)=\Gamma(\Omega_f)e^{a^\dagger(u)}\Phi_{\mathcal H}
 =e^{\hat a^\dagger(\Omega_f u)}\Gamma(\Omega_f)\Phi_{\mathcal H}
 =e^{\hat a^\dagger(\Omega_f u)}\Phi_{\mathscr H^+}=e(\Omega_f u)\,.
\end{equation*}
Recall the definition of the basic creation, annihilation and conservation processes in $\Gamma_s(\mathcal H)$
\begin{definition}[creation, annihilation and conservation processes]
\label{creation_annihilation_conservation_def}
In the definition of $\tilde{\mathcal H}$ let $\mathcal H_0$ be chosen
to be trivial i.e., $\mathcal H_0=\mathbb C$ so that $\tilde{\mathcal H}\equiv \Gamma_s(\mathcal H)$. 
Let $m$ be a $\xi_{\mathcal H}$-martingale and
define $(\xi,\mathbb C,\mathcal H)$-regular adapted processes
$A_m^\dagger:=\{A_m^\dagger(t)\mid t\geq 0\}$ and $A_m:=\{A_m(t)\mid
t\geq 0\}$ by
\begin{enumerate}
 \item $D(A_m^\dagger(t))=D(A_m(t))=\mathcal E(\mathcal H)$,
 \item $A_m^\dagger(t)e(u)=a^\dagger(m_t)e(u)$, $u\in\mathcal H$,
 \item $A_m(t)e(u)=a(m_t)e(u)$, $u\in\mathcal H$.
\end{enumerate}
The process $A^\dagger_m$ is called the \emph{creation process} and the process $A_m$ is
called the \emph{annihilation process} in $\Gamma_s(\mathcal H)$.
Let $\mathbf K\in\mathcal B(\mathcal H)$ be an operator such that 
$[\mathbf K,\xi_{\mathcal H}([0,t])]=0$ for $t\geq 0$ and denote $\mathbf K_t:=\xi_{\mathcal H}([0,t])\mathbf K$. Define the
$(\xi,\mathbb C,\mathcal H)$-regular adapted process $\Lambda_{\mathbf K}=\{\Lambda_{\mathbf K}(t)\mid t\geq 0\}$
in $\Gamma_s(\mathcal H)$ by
\begin{enumerate}
 \item $D(\Lambda_{\mathbf K}(t))=\mathcal E(\mathcal H)$,
 \item $\Lambda_{\mathbf H}(t)e(u)=\lambda(\mathbf H_t)e(u)$, $\forall u\in\mathcal H$.
\end{enumerate}
The process $\Lambda_{\mathbf K}$ is called the \emph{conservation process} in
$\Gamma_s(\mathcal H)$ associated with $\mathbf K$.\hfill$\square$
\end{definition}
The creation, annihilation and conservation processes in $\Gamma_s(\mathscr H^+(\mathbb R))$ are defined
as in Definition \ref{creation_annihilation_conservation_def} with obvious changes. We denote by 
$\hat A_{\hat m}=\{\hat A_{\hat m}(t)\mid t\geq 0\}$, $\hat A^\dagger_{\hat m}=\{\hat A^\dagger_{\hat m}(t)\mid t\geq 0\}$
and $\hat\Lambda_{\hat K}=\{\hat\Lambda_{\hat K(t)}\mid t\geq 0\}$ the creation, annihilation and conservation processes in
$\Gamma_s(\mathscr H^+(\mathbb R))$ with $\hat m$ a $\xi_{\mathscr H^+}$-martingale and 
$\hat K\in\mathcal B(\mathscr H^+(\mathbb R))$. Applying the transformation 
$\Gamma(\Omega_f)$ to the creation and annihilation processes in Definition \ref{creation_annihilation_conservation_def}
we obtain 
\begin{eqnarray}
 \Gamma(\Omega_f)A_m(t)e(u)=&\Gamma(\Omega_f)a(m_t)e(u)
 =\hat a(\hat m_t)\Gamma(\Omega_f)e(u)=\hat A_{\hat m}(t)\Gamma(\Omega_f)e(u)\,,\label{creation_intertwining_eqn}\\
 \Gamma(\Omega_f)A^\dagger_m(t)e(u)=&\Gamma(\Omega_f)a^\dagger(m_t)e(u)
 =\hat a^\dagger(\hat m_t)\Gamma(\Omega_f)e(u)=\hat A^\dagger_{\hat m}(t)\Gamma(\Omega_f)e(u)\,,
 \label{annihilation_intertwining_eqn}
 \end{eqnarray}
where according to Lemma \ref{martingale_mapping_lemma} $\hat m=\Omega_f m$ is a $\xi_{\mathscr H^+}$-martingale. 
Eqns. (\ref{creation_intertwining_eqn}), (\ref{annihilation_intertwining_eqn}) can be written in short form
\begin{equation}
\label{element_process_trans_A_eqn}
 \Gamma(\Omega_f)A_m=\hat A_{\hat m}\Gamma(\Omega_f),\qquad
 \Gamma(\Omega_f)A^\dagger_m=\hat A^\dagger_{\hat m}\Gamma(\Omega_f)\,.
\end{equation}
Let $\mathbf K\in\mathcal B(\mathcal H)$ be such that $[\mathbf K,\xi_{\mathcal H}(E)]=0$ for all $E\in\mathfrak B^+$
and define $\hat K$ according to Eq. (\ref{hat_K_definition_eqn}). Then by Lemma \ref{K_to_hat_K_lemma} $\hat K\in\mathcal B(\mathscr H^+)$ and
$[\hat K,\xi_{\mathscr H^+}(E)]=0$. Moreover, denoting $\mathbf K_t=\xi_{\mathcal H}([0,t])\mathbf K$, the same lemma, and in particular 
Eq. (\ref{K_measure_intertwine_eqn}), imply that 
\begin{equation*}
 \widehat{(K_t)}=\overline{(\Omega_f^*)^{-1}\mathbf K_t\Omega_f^*}=\overline{(\Omega_f^*)^{-1}\xi_{\mathcal H}([0,t])\mathbf K\Omega_f^*}
 =\xi_{\mathscr H^+}([0,t])\hat K=\hat K_t\,.
\end{equation*}
Therefore, applying $\Gamma(\Omega_f)$ to the conservation process $\Lambda_{\mathbf K}$ and using Definition \ref{mapping_algebra_def} we obtain
\begin{multline*}
 \Gamma(\Omega_f)\Lambda_{\mathbf K}(t)e(u)=\Gamma(\Omega_f)\lambda(\mathbf K_t)e(u)
 =\hat\lambda(\widehat{(K_t)})\Gamma(\Omega_f)e(u)=\\
 =\hat\lambda(\hat K_t)\Gamma(\Omega_f)e(u) =\Lambda_{\hat K}(t)\Gamma(\Omega_f)e(u)\,.
\end{multline*}
This equation can also be written in short form
\begin{equation}
\label{element_process_trans_B_eqn}
 \Gamma(\Omega_f)\Lambda_{\mathbf K}=\hat\Lambda_{\hat K}\Gamma(\Omega_f)\,.
\end{equation}
Eq. (\ref{element_process_trans_A_eqn}) and Eq. (\ref{element_process_trans_B_eqn}) provide the transformation properties
of the fundamental creation, annihilation and conservation processes under the mapping $\Gamma(\Omega_f)$. Since stochastic
integration, and subsequently the construction of quantum stochastic differential equations, is defined with respect to these basic
processes the transformation defined in Eq. (\ref{element_process_trans_A_eqn}) and Eq. (\ref{element_process_trans_B_eqn})
allows a mapping of stochastic processes defined with respect to $\tilde{\mathcal H}$ and the time observable $\mathbf T_F$ into stochastic
processes defined with respect to $\tilde{\mathscr H}^+$ and the time observable $\hat T_F$. The procedure for doing this is
demonstrated in the next example.
\subsection{Mappings of quantum stochastic processes}
\label{stoch_process_map_subsection}
\par We give an example of the mapping of stochastic processes induced by the map $\Omega_f$ through the transformation
$\Gamma(\Omega_f)$. Before doing that we need to complete the discussion of the previous subsection with one more step.
Suppose that $\mathcal K$ is a complex seperable Hilbert space and that $\xi$ is an $\mathbb R^+$-valued time observable
defined in $\mathcal K$. By this we mean that $\xi$ is the spectral measure of some self-adjoint operator $T$ with spectrum
$\mathbb R^+$. Suppose that $m$ and $m'$ are two $\xi$-martingales. Then there is a complex 
measure $\ll m,m' \gg$ in $\mathbb R^+$ satisfying \cite{Par2}
\begin{equation*}
 \ll m,m'\gg([0,t])=(m_t,m'_t)_{\mathcal K},\qquad \forall t\geq 0\,.
\end{equation*}
For the example given below we shall need the transformation properties of this complex measure under
the transformation $\Gamma(\Omega_f)$. Letting $\mathcal K=\mathcal H$ we recall that for every
exponential vector in $\Gamma_s(\mathcal H)$ (indeed this extends to every element of $\mathcal E(\mathcal H)$) 
\begin{equation*}
 [a(u),a^\dagger(v)]e(w)=(u,v)_{\mathcal H}e(w),\quad u,v,w\in\mathcal H\,.
\end{equation*}
Hence, if $m$, $m'$ are two $\xi_{\mathcal H}$-martingales we have
\begin{equation}
\label{a_a_dag_commutation_eqn}
 [a(m_t),a^\dagger(m'_t)]e(u)=(m_t,m'_t)_{\mathcal H}e(u)=\ll m,m' \gg([0,t])e(u)\,.
\end{equation}
Applying the transformation $\Gamma(\Omega_f)$ to Eq. (\ref{a_a_dag_commutation_eqn}) we obtain
\begin{multline*}
 \Gamma(\Omega_f)\left[\ll m,m' \gg([0,t])e(u)\right]=\Gamma(\Omega_f)[a(m_t),a^\dagger(m'_t)]e(u)
 =[\hat a(\hat m_t),\hat a^\dagger(\hat m'_t)]\Gamma(\Omega_f)e(u)=\\
 =(\hat m_t,\hat m'_t)_{\mathscr H^+}\Gamma(\Omega_f)e(u)=\ll \hat m,\hat m' \gg([0,t])\Gamma(\Omega_f)e(u)\,.
\end{multline*}
We can now give an example of the mapping of stochastic processes in the form of the application of the transformation
$\Gamma(\Omega_f)$ to an important class of quantum stochastic differential equations considered in \cite{HuP,Par2}. Let 
$\tilde{\mathcal H}$ and $\mathcal H_0$ be Hilbert spaces as in the left member of Eq. (\ref{tilde_H_tilde_H_plus_eqn}).
Let $O$ be a bounded operator in $\mathcal H_0$. We shall use the same notation for the extension $O=\{O(t)\mid t\geq 0\}$
of $O$ to a constant regular adapted process with respect to the triplet $(\xi_{\mathcal H},\mathcal H_0,\mathcal H)$
defined by the ampliation $O(t)=O\otimes\mathbf 1_{\Gamma_s(\mathcal H)}$. In \cite{HuP,Par2} we find the following theorem which is stated
here for the case of the Hilbert space $\mathcal H$ and the time observable $\mathbf T_F$:
\begin{theorem}
Let $L,S,H\in\mathcal B(\mathcal H_0)$ and assume that $S$ is unitary and $H$ is self-adjoint. Let $\mathbf P$ be
a projection in $\mathcal H$ commuting with the spectral measure $\xi_{\mathcal H}$ of the time observable
$\mathbf T_F$. Let $m$ be a $\xi_{\mathcal H}$-martingale such that $\mathbf P m_t=m_t$, $\forall t\geq 0$. Then there exists a
unique unitary operator-valued $(\xi_{\mathcal H},\mathcal H_0,\mathcal H)$-regular adapted process
$U=\{U(t)\mid t\geq 0\}$ satisfying
\begin{equation}
\label{physical_unitary_process_eqn}
 dU=\left[L\,dA^\dagger_m+(S-1)d\Lambda_{\mathbf P}-L^*S\,dA_m-(iH+\frac{1}{2}L^*L)d\ll m,m\gg\right]U,\ 
 U(0)=\mathbf 1_{\tilde{\mathcal H}}\,.
\end{equation}
Where $A^\dagger_m$, $A_m$ and $\Lambda_{\mathbf P}$ are the fundamental creation, annihilation and conservation processes in
$\Gamma_s(\mathcal H)$.\hfill$\square$
\end{theorem}
In order to be applicable to the stochastic process in Eq. (\ref{physical_unitary_process_eqn}) we first extend
$\Gamma(\Omega_f)$ to a mapping $\tilde\Gamma(\Omega_f): \mathcal B(\mathcal H_0)\otimes\mathscr W\mapsto
\mathcal B(\mathcal H_0)\otimes\hat{\mathscr W}$ via the relation
\begin{equation*}
 \tilde\Gamma(\Omega_f)(A\otimes B)=A\otimes\Gamma(\Omega_f)B,\qquad A\in\mathcal B(\mathcal H_0),\
 B\in\mathscr W\,.
\end{equation*}
It is now possible to extend $\tilde\Gamma(\Omega_f)$ to a transformation of quantum stochastic process solutions of
Eq. (\ref{physical_unitary_process_eqn}). First write a formal expression for the transformation of the stochastic process $U$
into a stochastic process $\hat U$ by the (extended) mapping $\tilde\Gamma(\Omega_f)$ in the form 
$\tilde\Gamma(\Omega_f)U=\hat U\tilde\Gamma(\Omega_f)$. Then, in order to define the process $\hat U$ and thus complete the
definition of the mapping $\tilde\Gamma(\Omega_f)$ apply the transformation $\tilde\Gamma(\Omega_f)$ to 
Eq. (\ref{physical_unitary_process_eqn}) and use the already known transformation properties of $L$, $S$, $H$, $\ll m,m\gg$
and $A_m$, $A^\dagger_m$, $\Lambda_{\mathbf P}$ to get
\begin{multline}
\label{hardy_unitary_process_eqn}
 d\left[\tilde\Gamma(\Omega_f)U\right]=\tilde\Gamma(\Omega_f)
 \left[L\,dA^\dagger_m+(S-1)d\Lambda_{\mathbf P}-L^*S\,dA_m-(iH+\frac{1}{2}L^*L)d\ll m,m\gg\right]U=\\
 = \left[L\,d\hat A^\dagger_{\hat m}+(S-1)d\hat\Lambda_{\hat P}-L^*S\,d\hat A_{\hat m}
 -(iH+\frac{1}{2}L^*L)d\ll \hat m,\hat m\gg\right]\tilde\Gamma(\Omega_f)U\,,
\end{multline}
where in Eq. (\ref{hardy_unitary_process_eqn}) $L$, $S$ and $H$ stand for the constant stochastic processes
$L(t)=L\otimes\mathbf 1_{\Gamma_s(\mathscr H^+)}$, $S(t)=S\otimes\mathbf 1_{\Gamma_s(\mathscr H^+)}$
and $H(t)=H\otimes\mathbf 1_{\Gamma_s(\mathscr H^+)}$ in $\tilde{\mathscr H}^+$. Of course, 
Eq. (\ref{hardy_unitary_process_eqn}) is still formal. However, the definition of $\hat U$ is clear, i.e., 
$\hat U$ is naturally defined as the solution of the quantum stochastic differential equation 
\begin{equation}
\label{hardy_unitary_process_A_eqn}
 d\hat U=\left[L\,d\hat A^\dagger_{\hat m}+(S-1)d\hat\Lambda_{\hat P}-L^*S\,d\hat A_{\hat m}
 -(iH+\frac{1}{2}L^*L)d\ll \hat m,\hat m\gg\right]\hat U,\ \ \hat U(0)=\mathbf 1_{\Gamma_s(\tilde{\mathscr H}^+)}\,.
\end{equation}
The transition from Eq. (\ref{physical_unitary_process_eqn}) to Eq. (\ref{hardy_unitary_process_eqn}) can then be written 
\begin{multline*}
 \tilde\Gamma(\Omega_f)dU
 =\tilde\Gamma(\Omega_f)\left[L\,dA^\dagger_m+(S-1)d\Lambda_P-L^*S\,dA_m-(iH+\frac{1}{2}L^*L)d\ll m,m\gg\right]U\\
 \Downarrow\\
 d\hat U \;\tilde\Gamma(\Omega_f)=\left[L\,d\hat A^\dagger_{\hat m}+(S-1)d\hat\Lambda_{\hat P}-L^*S\,d\hat A_{\hat m}
 -(iH+\frac{1}{2}L^*L)d\ll \hat m,\hat m\gg\right]\hat U \;\tilde\Gamma(\Omega_f)\,.\\ 
\end{multline*}
The transformation $\tilde\Gamma(\Omega_f)$ constructed as above is well defined on solutions $U$ 
of Eq. (\ref{physical_unitary_process_eqn}) and is given by the intertwining relation
\begin{equation}
\label{stochastic_process_intertwining_eqn}
 \tilde\Gamma(\Omega_f)U=\hat U\tilde\Gamma(\Omega_f)
\end{equation}
where $\hat U$ is a solution of Eq. (\ref{hardy_unitary_process_A_eqn}). In particular, the transformation of the initial state
$U(0)$ in Eq. (\ref{physical_unitary_process_eqn}) is given by
\begin{equation*}
 \tilde\Gamma(\Omega_f)U(0)= \tilde\Gamma(\Omega_f)\mathbf 1_{\tilde{\mathcal H}}
 =\mathbf 1_{\tilde{\mathscr H}^+}\tilde\Gamma(\Omega_f)=\hat U(0)\tilde\Gamma(\Omega_f)
\end{equation*}
where $\hat U(0)=\mathbf 1_{\tilde{\mathscr H}^+}$ is the initial value for the process $\hat U$.
\par Eq. (\ref{stochastic_process_intertwining_eqn}) is an analogue, for the class of quantum stochastic differential equations
considered here, of the fundamental intertwining relation in Eq. (\ref{basic_intertwining_rel_eqn}). We observe that the stochastic process $U$ is defined
with respect to the spectral measure $\xi_{\mathcal H}$ of the time observable $\mathbf T_F$, the stochastic process
$\hat U$ is defined with respect to the spectral measure $\xi_{\mathscr H^+}$ of the observable $\hat T_F$ and the transformation
$\tilde\Gamma(\Omega_f)$ of stochastic processes is induced by the mapping 
$\Omega_f:\mathcal H\mapsto\mathscr H^+(\mathbb R)$. It should be remarked at this point that following
a procedure very similar to the one presented in this section it is possible to define an induced transformation 
$\tilde\Gamma(\Omega_f^*)$ mapping stochastic differential equations defined in $\tilde{\mathscr H}^+$ into stochastic
differential equations defined in $\tilde{\mathcal H}$.
\section{Summary}
\label{summary_sec}
\par Time observables $\mathbf T_F$  and $\mathbf T_B$ for forward and backward quantum evolution were introduced in 
Section \ref{the_time_observables_sec} above under the assumption that the quantum system under consideration is
described by a complex seperable 
Hilbert space $\mathcal H$ and the generator of evolution is a self-adjoint Hamiltonian $\mathbf H$ on $\mathcal H$ satifying
$\sigma(\mathbf H)=\sigma_{ac}(\mathbf H)=\mathbb R^+$. It was shown in Section \ref{the_time_observables_sec} 
that $\mathbf T_F$ and $\mathbf T_B$
are positive, self-adjoint, semi-bounded operators in $\mathcal H$. The characterization of $\mathbf T_F$ as a forward time
observable emerges from the fact, proved in Theorem \ref{time_observable_thm} in Section \ref{the_time_observables_sec}, 
that if the quantum evolution is applied in the 
forward direction (i.e., for positive times) to an initial state $g\in\mathcal H$ supported in a finite interval $\Delta$ in the spectrum of
$\mathbf T_F$, then the evolved state $g(t)=\mathbf U(t)g=\exp[-i\mathbf H t]g$, $t\geq 0$ necessarily flows to higher parts of 
the spectrum of $\mathbf T_F$ as $t$ increases. Indeed for any finite interval $\Delta\in\sigma(\mathbf T_F)$
the norm of the projection of $g(t)$ on the subspace of $\mathcal H$ corresponding to $\Delta$ by the spectral theorem (applied to
$\mathbf T_F$) goes to zero as $t$ goes to infinity. An analogous result holds for $\mathbf T_B$ for backward time evolution.
\par The basic mechanism enabling the definition of the time observable $\mathbf T_F$ involves a central
ingredient of the semigroup decomposition formalism in the form of the fundamental intertwining relation appearing in 
Eq. (\ref{basic_intertwining_rel_eqn}). The fact that the characterization of $\mathbf T_F$ as a forward time observable is achieved
through this intertwining relation, valid only for forward evolution, whereas the intertwining relation in 
Eq. (\ref{back_basic_intertwine_eqn}), leading to the characterization of $\mathbf T_B$ as a backward time observable, is valid
only for backward evolution, displays a built in time asymmetry in the theory. 
In Section \ref{t_observe_stochastic_process_sec} 
the discussion of this time asymmetry is opened up a bit further. it is shown there that, beyond
its applicability to future directed Schr\"odinger type evolution, the operator $\mathbf T_F$ can, in fact, be used as a time
observable in the construction of more general types of quantum processes clearly exhibiting future directed time evolution.
Specifically, $\mathbf T_F$ and the corresponding Hardy space time observable $\hat T_F$ are used in the construction of
quantum stochastic differential equations the solutions of which are (quantum) stochastic processes shown to satisfy an
intertwining relation analogous to Eq. (\ref{basic_intertwining_rel_eqn}). Moreover, the map intertwining these quantum stochastic
processes is, in fact, induced by the map $\Omega_f$ appearing in Eq. (\ref{basic_intertwining_rel_eqn}). Of course, the whole
discussion can be repeated for backward time evolution using the operators $\mathbf T_B$ and $\hat T_B$.
\par Many questions are, of course, left open regarding the nature of the time observables and their appications. Here we mention
briefly just a few. A first important question is whether the restriction on the spectrum of the Hamiltonian $\mathbf H$ put at the
begining of the paper can be relaxed in such a way that meaningful time observables can still be defined.
In addition, since $\mathbf T_F$ and $\mathbf T_B$ are operators in the physical space $\mathcal H$, one would like, if possible,
to obtain expressions for them directly in terms of physical space variables without need for mappings to Hardy spaces. In this
case what are the relations of the time ovservables to other observables of the physical system such as the Hamiltonian,
momentum, position etc.\ ? In the context of irreversible quantum evolutions and the discusstion in
Section \ref{t_observe_stochastic_process_sec}
concerning quantum stochastic processes, one of the questions immediately arising is whether the intertwining relation in 
Eq. (\ref{stochastic_process_intertwining_eqn} ) can be shown to be more than just merely analogous to the one in
Eq. (\ref{basic_intertwining_rel_eqn}) i.e., is it
possible to show, for example, that Eq. (\ref{basic_intertwining_rel_eqn}) can be recovered in some sense from 
Eq. (\ref{stochastic_process_intertwining_eqn}). These and related problems will be addressed elsewhere.
\section{Appendix A}
\par In this appendix we consider the properties of
$\mathbf T_F^{-1}$ and $\hat T_F^{-1}$ as Hilbert space contractions on
$\mathcal H$ and $\mathscr H^+(\mathbb R)$ respectively. We recall
that the \emph{defect operators} for a contraction $T: \mathcal
H\mapsto\mathcal H'$ and its adjoint $T^*:\mathcal H'\mapsto\mathcal
H$ are given by
\begin{equation*}
 D_T:=(1-T^*T)^{1/2},\qquad D_{T^*}:=(1-TT^*)^{1/2}
\end{equation*}
and the defect subspaces $\mathscr D_T$ and $\mathscr D_{T^*}$ are defined
by
\begin{equation*}
 \mathscr D_T:=\overline{D_T\mathcal H},\qquad \mathscr
 D_{T^*}=\overline{D_{T^*}\mathcal H'}\,.
\end{equation*}
Hence for $\mathbf T_F$ and $\hat T_F^{-1}$
\begin{equation}
\label{tf_defect_opr_eqn}
 D_{\mathbf T_F^{-1}}=(1-(\mathbf T_F^{-1})^2)^{1/2}=(1-(\Omega_f^*\Omega_f)^2)^{1/2},\qquad
 D_{\hat T_F^{-1}}=(1-(\hat T_F^{-1})^2)^{1/2}=(1-(\Omega_f\Omega_f^*)^2)^{1/2}
\end{equation}
and for $\Omega_f$ and $\Omega_f^*$ we have
\begin{equation}
\label{omega_defect_opr_eqn}
 D_{\Omega_f}=(1-\Omega_f^*\Omega_f)^{1/2}=(1-\mathbf T_F^{-1})^{1/2},\qquad
 D_{\Omega_f^*}=(1-\Omega_f\Omega_f^*)^{1/2}=(1-\hat T_F^{-1})^{1/2}\,.
\end{equation}
We note that $(1+\mathbf T_F^{-1})^{1/2}\mathcal H=\mathcal H$ and 
$(1+\hat T_F^{-1})^{1/2}\mathscr H^+(\mathbb R)=\mathscr H^+(\mathbb R)$. Using Eq. (\ref{tf_defect_opr_eqn}) and
Eq. (\ref{omega_defect_opr_eqn}) we obtain
\begin{equation*}
 D_{\mathbf T_F^{-1}}=D_{\Omega_f}(1+\mathbf T_F)^{1/2},\quad
 D_{\hat T_F^{-1}}=D_{\Omega_f^*}(1+\hat T_F)^{1/2}\,.
\end{equation*}
Hence we get that
\begin{eqnarray}
\label{tf_eql_defect_omega_eqn}
 \mathscr D_{\mathbf T_F^{-1}}&=&\overline{D_{\mathbf T_F^{-1}}\mathcal H}=\overline{D_{\Omega_f}\mathcal H}
 =\mathscr D_{\Omega_f},\\
 \mathscr D_{\hat T_F^{-1}}&=&\overline{D_{\hat T_F^{-1}}\mathscr H^+(\mathbb R)}
 =\overline{D_{\Omega_f^*}\mathscr H^+(\mathbb R)}
 =\mathscr D_{\Omega_f^*}\,.
\end{eqnarray}
In addition we have the relations
\begin{equation*}
 \mathscr D_{T^*}=\overline{T\mathscr D_T}\oplus\text{Ker}\,(T^*),\qquad
 \mathscr D_{T}=\overline{T^*\mathscr D_{T^*}}\oplus\text{Ker}\,(T)\,.
\end{equation*}
so that by the fact that $\Omega_f$ and $\Omega_f^*$ are both
injective we have
\begin{equation*}
 \mathscr D_{\Omega_f^*}=\overline{\Omega_f\mathscr D_{\Omega_f}},\qquad
 \mathscr D_{\Omega_f}=\overline{\Omega_f^*\mathscr D_{\Omega_f^*}}\,.
\end{equation*}
Hence
\begin{equation*}
 \Omega_f\mathscr D_{\Omega_f}\subset\mathscr D_{\Omega_f^*},\qquad
 \Omega_f^*\mathscr D_{\Omega_f^*}\subset\mathscr D_{\Omega_f}\,.
\end{equation*}
Eq. (\ref{tf_eql_defect_omega_eqn}) then implies that
\begin{equation}
\label{defect_subspace_tf_hat_tf_eqn}
 \Omega_f\mathscr D_{\mathbf T_F^{-1}}\subset\mathscr D_{\hat T_F^{-1}},\qquad
 \Omega_f^*\mathscr D_{\hat T_F^{-1}}\subset\mathscr D_{\mathbf T_F^{-1}}\,.
\end{equation}
In fact, in Eq. (\ref{defect_subspace_tf_hat_tf_eqn}) the left hand
members are dense in the right hand members.
%
%
%
%
The \emph{charateristic function} $\Theta_{\mathbf T_F^{-1}}: \mathscr
D_{\mathbf T_F^{-1}}\mapsto\mathscr D_{\mathbf T_F^{-1}}$ of $T_F^{-1}$ is given by
\begin{equation}
\label{tf_characteristic_func_eqn}
 \Theta_{\mathbf T_F^{-1}}(\lambda)
  =\left(-\mathbf T_F^{-1}+\lambda D_{\mathbf T_F^{-1}}(1-\lambda\mathbf T_F^{-1})^{-1} D_{T_F^{-1}}\right)
 \big\vert\mathscr D_{\mathbf T_F^{-1}}\,.
\end{equation}
Hence, with the help of Eq. (\ref{hat_rtf_rtf_eqn} ), we can write 
\begin{multline}
\label{tf_characteristic_func_A_eqn}
 \Theta_{\mathbf T_F^{-1}}(\lambda)
  =\bigg[\lambda(1-\lambda\Omega_f^*\Omega_f)^{-1}
  -\left(\Omega_f^*\Omega_f
  +\lambda (\Omega_f^*\Omega_f)^2(1-\lambda\Omega_f^*\Omega_f)^{-1}\right)\bigg]\bigg\vert\mathscr  
  D_{\mathbf T_F^{-1}}=\\
  =\bigg[\lambda(1-\lambda\Omega_f^*\Omega_f)^{-1}
  -\Omega_f^*\left(1 
  +\lambda\Omega_f(1-\lambda\Omega_f^*\Omega_f)^{-1}\Omega_f^*\right)\Omega_f\bigg]\bigg\vert\mathscr  
  D_{\mathbf T_F^{-1}}=\\
  =\bigg[R_{\mathbf T_F^{-1}}(\lambda^{-1})
  -\lambda^{-1}\Omega_f^*R_{\hat T_F^{-1}}(\lambda^{-1})
  \Omega_f\bigg]\bigg\vert\mathscr D_{\mathbf T_F^{-1}}\,.\\
\end{multline}
Exchanging $\Omega_f$ and $\Omega_f^*$ in
Eq. (\ref{tf_characteristic_func_A_eqn}) we get the characteristic
function of $\hat T_F^{-1}$
\begin{equation}
\label{hat_tf_characteristic_func_eqn}
 \Theta_{\hat T_F^{-1}}(\lambda)
 =\bigg[R_{\hat T_F^{-1}}(\lambda^{-1})-\lambda^{-1}\Omega_f R_{\mathbf T_F^{-1}}(\lambda^{-1})
 \Omega_f^*\bigg]\bigg\vert\mathscr D_{\hat T_F^{-1}}\,.
\end{equation}
Multiplying Eq. (\ref{hat_tf_characteristic_func_eqn}) from the left
by $\Omega_f^*$ and takeing notice of the fact that the identity
$\Omega_f\mathbf T_F^{-1}=\hat T_F^{-1}\Omega_f^*$ imply that
$\Omega_f R_{\mathbf T_F^{-1}}(z)=R_{\hat T_F^{-1}}(z)\Omega_f$ and 
$\Omega_f^* R_{\hat T_F^{-1}}(z)=R_{\mathbf T_F^{-1}}(z)\Omega_f^*$ we
have
\begin{equation}
\label{omega_star_mult_eqn}
 \Omega_f^*\Theta_{\hat T_F^{-1}}(\lambda)
 =\bigg[R_{\mathbf T_F^{-1}}(\lambda^{-1})-\lambda^{-1}\Omega_f^* R_{\hat T_F^{-1}}(\lambda^{-1})
  \Omega_f\bigg]\Omega_f^*\bigg\vert\mathscr D_{\hat T_F^{-1}}\,.
\end{equation}
We now compare Eq. (\ref{omega_star_mult_eqn}) and
Eq. (\ref{tf_characteristic_func_A_eqn}) and take into account
Eq. (\ref{defect_subspace_tf_hat_tf_eqn}) in order to obtain
\begin{equation*}
 \Omega_f^*\Theta_{\hat
   T_F^{-1}}(\lambda)=\Theta_{\mathbf T_F^{-1}}(\lambda)\Omega_f^*\big\vert\mathscr D_{\hat T_F^{-1}}\,.
\end{equation*}
Following a similar procedure we find also that
\begin{equation*}
 \Omega_f \Theta_{\mathbf T_f^{-1}}(\lambda)=\Theta_{\hat T_F^{-1}}(\lambda)\Omega_f \big\vert \mathscr D_{\mathbf T_F^{-1}}\,.
\end{equation*}
Thus we arrived at the following proposition
\begin{proposition}
\label{char_func_transform_prop}
Let $\Theta_{\mathbf T_F^{-1}}(z)$ and $\Theta_{\hat T_F^{-1}}(z)$ be the
characteristic and $\mathscr D_{\mathbf T_F^{-1}}$, $\mathscr D_{\hat T_F^{-1}}$ be the defect
subspaces of $\mathbf T_F^{-1}$ and $\hat T_F^{-1}$ respectively. Then we have
\begin{equation*}
 \Omega_f^*\Theta_{\hat
   T_F^{-1}}(z)=\Theta_{T_F^{-1}}(z)\Omega_f^*\big\vert\mathscr D_{\hat T_F^{-1}}\,.
\end{equation*}
\begin{equation*}
 \Omega_f \Theta_{\mathbf T_f^{-1}}(\lambda)=\Theta_{\hat T_F^{-1}}(\lambda)\Omega_f \big\vert \mathscr D_{\mathbf T_F^{-1}}\,.
\end{equation*}
\hfill$\square$
\end{proposition}
The equality of spectrums is an immediate corollary of the above proposition:
\begin{corollary}
The spectrums of $\mathbf T_F^{-1}$ and $\hat T_F^{-1}$ satisfy
$\sigma(\mathbf T_F^{-1})=\sigma(\hat T_F^{-1})$.\hfill$\square$
\end{corollary}
\par{\bf Proof:} 
\smallskip
\par We can use either of the two equations in Proposition \ref{char_func_transform_prop}. Using the first equation
the corollary immediately follows from the fact that $\text{Ker}\,(\Omega_f^*)=\{0\}$, the fact that $\Omega_f^*\mathscr
D_{\hat T_F^{-1}}$ is dense in $\mathscr D_{\mathbf T_F^{-1}}$ and from a theorem
of Sz.-Nagy and C. Foias (see \cite{SzNF}, Chap. VI, Sec. 4) stating that the
characteristic function of a contraction $T$ determines uniquely the
spectrum of $T$.\hfill$\blacksquare$
\par\bigskip\
\par\centerline{\Large\bf Acknowledgments}
\smallskip
\par I wish to thank Prof. L.P. Horwitz for useful discussions during the preparation of this paper. 
Research supported by ISF under Grant No. 1282/05 and by the Center for Avanced Studies in 
Mathematics at Ben-Gurion University. 
\par\bigskip\

\end{document}